\documentclass[aps,pre,twocolumn, groupaddress]{revtex4-2}

\usepackage{amsmath}
\usepackage{amssymb}
\usepackage{amsthm}
\usepackage[hidelinks]{hyperref}
\usepackage{graphicx}
\usepackage{stmaryrd}

\graphicspath{{fig/}}
\usepackage{xcolor}

\theoremstyle{plain} 

\DeclareMathOperator{\Diver}{Div}
\DeclareMathOperator{\Grad}{Grad}

\DeclareMathOperator{\tr}{tr}

\usepackage{mathtools}
\usepackage{pgfplots}
\pgfplotsset{/pgf/number format/use comma,compat=newest}

\renewcommand\epsilon{\varepsilon}

\newcommand{\vect}[1]{\mathbf{#1}}
\newcommand{\tens}[1]{\mathsf{#1}}

\hypersetup{%
    pdfborder = {0 0 0}
}
\begin{document}


\title{Active elasticity drives the formation of periodic beading in damaged axons}


\author{Davide Riccobelli}
\email[]{davide.riccobelli@polimi.it}
\affiliation{MOX -- Dipartimento di Matematica, Politecnico di Milano, Piazza Leonardo da Vinci 32, 20133 Milano, Italy}


\date{\today}

\begin{abstract}
In several pathological conditions, such as coronavirus infections, multiple sclerosis, Alzheimer's and Parkinson's diseases, the physiological shape of axons is altered and a periodic sequence of bulges appears. Experimental evidences suggest that such morphological changes are caused by the disruption of the microtubules composing the cytoskeleton of the axon. In this paper, we develop a mathematical model of damaged axons based on the theory of continuum mechanics and nonlinear elasticity. The axon is described as a cylinder composed of an inner passive part, called axoplasm, and an outer active cortex, composed mainly of F-actin and able to contract thanks to myosin-II motors. Through a linear stability analysis we show that, as the shear modulus of the axoplasm diminishes due to the disruption of the cytoskeleton, the active contraction of the cortex makes the cylindrical configuration unstable to axisymmetric perturbations, leading to a beading pattern. Finally, the non-linear evolution of the bifurcated branches is investigated through finite element simulations.
\end{abstract}


\maketitle


\section{Introduction}

\begin{figure}[b!]
\includegraphics[width=\columnwidth]{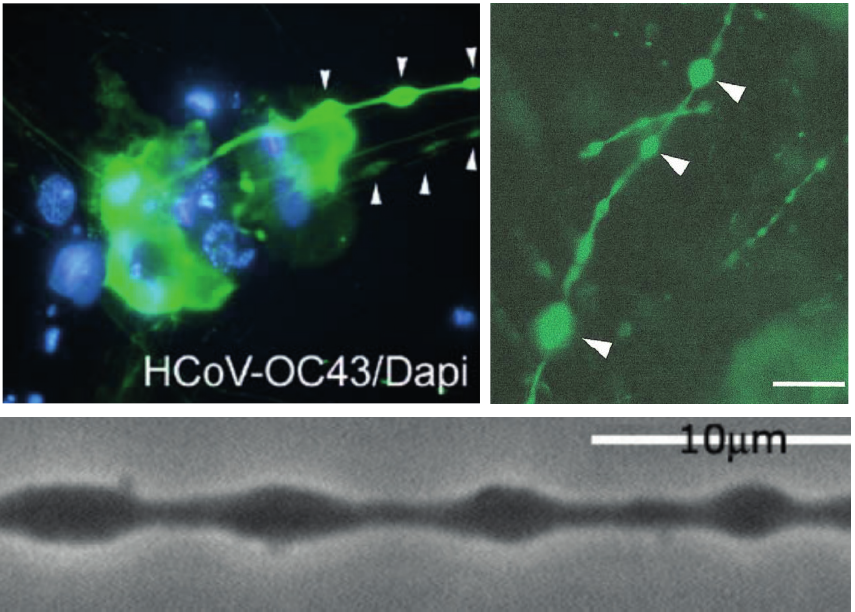}
\caption{(Top left) Periodic swellings in a rat axon induced by human coronavirus OC43, adapted from \cite{Jacomy_2006}. (Top, right) Axonal beading induced by oxidative stress (40 mM of H$_2$O$_2$), adapted from \cite{Roediger_2003}. (Bottom) Beading of a PC12 neurite after the exposure to $10\,\mu\mathrm{g}/\mathrm{ml}$ of nocodazole, adapted from \cite{Datar_2019}.}
\label{fig:sperimentali}
\end{figure}

The current pandemic of Sars-Cov-2 is raising growing concerns for its effects on the central nervous system. During the acute stage, signs of delirium, post-traumatic stress, depression, encephalitis, and neurocognitive disorders have been reported in patients affected by COVID-19 \cite{Beach_2020, Troyer_2020,ELLUL2020767}. Other coronavirus diseases, such as the SARS and the MERS, can cause similar symptoms \cite{Rogers_2020}. Experiments on mice have shown that human coronaviruses can attack the central nervous system, causing cytopathic effects on neurons \cite{Jacomy_2006}, namely the cells of the nervous tissue. They are composed of the soma, that is the central part containing the nucleus and the organelles, the dendrites, and a single axon. Axons and dendrites are structures which transmit electrochemical signals to and from the soma, respectively. In particular, the axon is composed of a long cylindrical filament, called axonal shaft, which can bifurcate into many branches at its end, called telodendria. In the experiments of Jacomy and co-workers \cite{Jacomy_2006}, the human coronavirus OC43 triggers the formation of a periodic peristaltic pattern along the axonal shaft (see Figure~\ref{fig:sperimentali}).
Such a morphological change is not an exclusive manifestation of coronavirus infections. Similar periodic swellings have been observed in axons affected by other pathologies, such as multiple sclerosis \cite{nikic2011reversible}, early stages of the Alzheimer's \cite{Stokin1282} and Parkinson's  disease \cite{tagliaferro2016retrograde}, and in response to traumatic stretch injuries \cite{bain2000tissue}. The formation of periodic bulges in the axon seems to diminish or even to inhibit its ability of transmitting electrical signals \cite{Kolaric_2013}.

In particular, it is believed that oxidative stress is implicated in the genesis of several neurodegenerative pathologies, such as the Alzheimer's disease \cite{nunomura2006involvement} and through \emph{in vitro} experiments it has been observed the formation of swellings along the axonal shaft after exposure to hydrogen peroxide (see Figure~\ref{fig:sperimentali}). In this case, the accumulation of $\beta$-tubulin III indicates that microtubules have been disrupted \cite{Roediger_2003}. In healthy neurons, microtubules are binded together composing the cytoskeleton, which is one of the main constituents of the inner region of the axonal shaft, called axoplasm. The axoplasm is surrounded by a cortex, mainly composed of F-actin filaments and myosin motors.
Furthermore, many neurodegenerative pathologies (such as the Alzheimer's and the Pick's diseases) are characterized by the presence of misfolded tau proteins which can destabilize microtubules and disrupt their network, reducing the elastic modulus of the axoplasm \cite{Woerman_2016,van_den_Bedem_2015,de_Rooij_2018}.
The importance of microtubule network integrity has been confirmed by other experiments: after being exposed to nocodazole, a potent microtubule depolymerizer, axons undergo a shape transition, exhibiting a periodic beading pattern along the axonal shaft and, simultaneously, the number of microtubules significantly decreases \cite{He_2002,Datar_2019}.

On these grounds, it has been suggested that axonal beading may be the result of a mechanical instability triggered by the coupling of the active contractility of the actin cortex and microtubule depletion \cite{Datar_2019}. According to Datar and co-workers, this is reminiscent of the elastic analogue of the Rayleigh-Plateau instability, where an elastic cylinder can be destabilized by the presence of surface tension \cite{Mora_2010}. In the case of an axon, the cylinder represents the axoplasm while the contractility of the F-actin  cortex is modeled as the action of the surface tension. From a linear analysis, it is possible to prove that the critical wavelength of the elastic Rayleigh-Plateau instability is infinite \cite{Mora_2010,Taffetani_2015}. This instability shares many similarities with phase-transition phenomena \cite{Xuan_2017} and it has been recently shown that the resulting buckling configuration is characterized by a single localized swelling \cite{Lestringant_2020,Giudici_2020,Fu_2021} rather than a periodic beading as in damaged axons. 
This suggests that some other mechanism is involved in the buckling of axons.
In this respect, it is important to take into account some aspects. First, the elasticity of the axoplasm seems to be fundamental in maintaining the shape of the axon \cite{Pullarkat_2006,Datar_2019}. Second, the  cortex is composed of a network of F-actin filaments connected together by myosin motors and spectrin: it may be oversimplifying to model it as a surface tension acting on the axoplasm, neglecting its elasticity.
Finally, the thickness of the actin cortex is about $80$-$100\,\mathrm{nm}$ \cite{Letourneau_2009} which is not negligible compared with the radius of a human axon (about $300$-$500\,\mathrm{nm}$ \cite{Liewald_2014}).

In this paper, we show how axonal beading can be explained as the result of a purely elastic instability using a simple mathematical model based on continuum mechanics. The model is constructed in Section~\ref{sec:model}, while the stability of the cylindrical shape of the axon is investigated through a linear analysis in Section~\ref{sec:lin_stab}. In Section~\ref{sec:num} we report the outcomes of the numerical post-buckling analysis. Finally, the main results are summarized in Section~\ref{sec:rem} together with some concluding remarks.

\section{The model}
\label{sec:model}
We denote by $\Omega_0$ the reference domain of the axonal shaft, which is modeled as a cylinder of radius $R_o$. Let $\vect{X}\in\Omega_0$ be the material position vector, whose cylindrical coordinates are $(R,\,\Theta,\,Z)$.
Within this reference configuration, we identify two subregions
\begin{align*}
&\Omega_0^i = \left\{\vect{X}\in\Omega_0\;|\;0\leq R<R_i\right\},\\
&\Omega_0^o=  \left\{\vect{X}\in\Omega_0\;|\;R_i\leq R<R_o\right\},
\end{align*}
which are the subdomains representing the axoplasm and the peripheral region occupied by the F-actin cortex, respectively.

\begin{figure}[b!]
\includegraphics[width=\columnwidth]{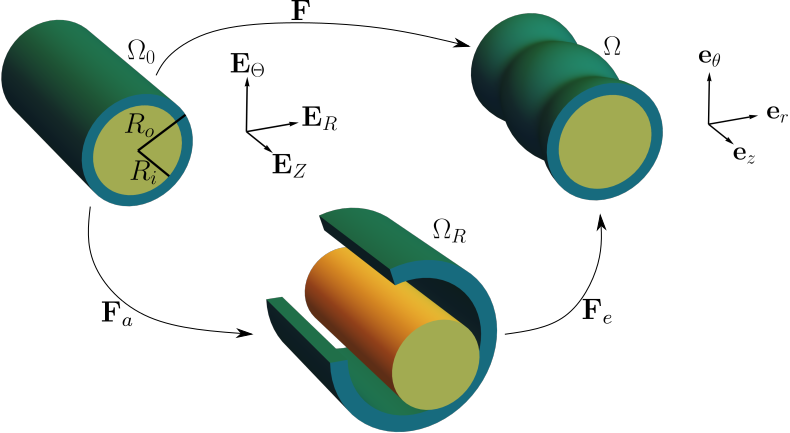}
\caption{Representation of the reference configuration $\Omega_0$, the relaxed state $\Omega_R$, and the actual configuration $\Omega$ according to the active strain theory.}
\label{fig:conf}
\end{figure}

Let $\vect{x}=\boldsymbol\chi(\vect{X}) = \vect{X} + \vect{u}(\vect{X})$ be the actual position vector, where $(r,\,\theta,\,z)$ are the actual cylindrical coordinates, while $\boldsymbol\chi$ and $\vect{u}$ are the deformation and the displacement fields, respectively. We denote by $\Omega=\boldsymbol{\chi}(\Omega_0)$ the actual configuration and let $\tens{F}=\Grad\vect{\chi}$ be the deformation gradient. The active contraction of the cortex is modeled through the so called \emph{active strain} approach \cite{Kondaurov_1987,Ambrosi_2011,Giantesio_2018}. In particular, a multiplicative decomposition of the deformation gradient is assumed, i.e.
\[
\tens{F}=\tens{F}_e\tens{F}_a,
\]
where $\tens{F}_a$ is the active strain tensor describing the micro-structural reorganization caused by the cortex contractility and maps the reference configurations to the relaxed state $\Omega_R$, while $\tens{F}_e$ accounts for the local elastic distortion, see Figure~\ref{fig:conf}. 
We remark that the tensor $\tens{F}_a$ represents a remodeling of the material: no mass is added or subtracted during the contraction of the cortex, so that the mass density remains constant. Mathematically, this can be enforced by requiring that $\det\tens{F}_a=1$ \cite{Epstein_2012}.
Furthermore, since actin filaments are mainly directed along the axial and the hoop direction \cite{Costa_2018}, a possible choice for the active strain tensor is given by
\begin{equation}
\label{eq:Fa}
\tens{F}_a = \left\{
\begin{aligned}
&\tens{I} &&\text{in }\Omega_0^i,\\
&\frac{1}{\lambda_a^2}\vect{E}_R\otimes\vect{E}_R+\lambda_a(\tens{I}-\vect{E}_R\otimes\vect{E}_R) &&\text{in }\Omega_0^o,
\end{aligned}
\right.
\end{equation} 
where $\tens{I}$ is the identity tensor, $\lambda_a\in(0,1]$ is the active stretch and  $(\vect{E}_R,\,\vect{E}_\Theta,\,\vect{E}_Z)$ is the cylindrical vector basis in the reference configuration. According to \eqref{eq:Fa}, the axoplasm is passive while the actin cortex contracts isotropically along the directions orthogonal to $\vect{E}_R$.

Furthermore, since the axon is mainly composed of water \cite{LoPachin_1991}, it is reasonable to assume that it is incompressible. In particular, we describe it as an elastic body composed of an incompressible neo-Hookean material, whose strain energy is given by
\begin{equation}
\label{eq:energy}
\Psi= \frac{\mu_\beta}{2}\left(\tr(\tens{F}_e^T\tens{F}_e) -3\right)\qquad\beta=i,\,o,
\end{equation}
where $\mu_i$ and $\mu_o$ are the shear moduli of the axoplasm and of the cortex, respectively.
The balance of linear momentum in quasi-static conditions is given by 
\begin{equation}
\label{eq:divP}
\Diver\tens{P} = \vect{0},
\end{equation}
where $\Diver$ is the divergence operator and $\tens{P}$ is the nominal stress tensor. Exploiting the Clausius-Duhem inequality, we get
\begin{equation}
\label{eq:P}
\tens{P}=\frac{\partial \Psi}{\partial\tens{F}} - p \tens{F}^{-1},
\end{equation}
where $p$ is the Lagrange multiplier enforcing the incompressibility constraint 
\begin{equation}
\label{eq:inc}
\det\tens{F}=1.
\end{equation}
Furthermore, the external surface is assumed to be stress free, i.e.
\begin{equation}
\label{eq:BC}
\tens{P}^T\vect{E}_R=\vect{0},
\end{equation}
and the total length of the axonal shaft is kept fixed during the deformation.
Finally, the continuity of the displacement and of the normal stress at the interface between the cortex and the inner core is enforced, namely \begin{equation}
\label{eq:IC}
\left\{
\begin{aligned}
&\llbracket\vect{u}\rrbracket=\vect{0} &&\text{at }R=R_i,\\
&\llbracket\tens{P}^T\vect{E}_R\rrbracket=\vect{0}&&\text{at }R=R_i,
\end{aligned}
\right.
\end{equation}
where $\llbracket\cdot\rrbracket$ denotes the jump operator.

It is straightforward to show that the reference configuration is in mechanical equilibrium. In fact, the balance of the linear momentum reduces to 
\begin{equation}
\label{eq:linear_balance}
\frac{d P_{RR}}{d R} + \frac{P_{RR}-P_{\Theta\Theta}}{R}=0.
\end{equation}
Since in this case $\tens{F}=\tens{I}$, using \eqref{eq:energy} and \eqref{eq:P} the nominal stress tensor can be written as
\begin{equation}
\label{eq:nominal}
\tens{P} = \left\{
\begin{aligned}
&\mu_i\tens{I} - p \tens{I} &&0\leq R<R_i,\\
&\mu_o\tens{F}_a^{-1}\tens{F}_a^{-T} - p \tens{I} &&R_i\leq R <R_o.\\
\end{aligned}
\right.
\end{equation}
Substituting \eqref{eq:nominal} into \eqref{eq:linear_balance}, one obtains
\begin{equation}
\label{eq:pres}
\left\{
\begin{aligned}
&p(R) = C_1 &&0\leq R<R_i,\\
&p(R)=\frac{\mu_o \left(\lambda_a^6-1\right)\log (R)}{\lambda_a^2}+C_2 &&R_i\leq R <R_o,
\end{aligned}
\right.
\end{equation}
where $C_1$ and $C_2$ are constants that are to be fixed by enforcing the boundary and the interface conditions \eqref{eq:BC}-\eqref{eq:IC}. More explicitly, from \eqref{eq:BC} it is possible to get an analytical expression for $C_2$
\[
C_2 = \frac{\mu_o \left(\lambda_a^6-\lambda_a^6 \log (R_o)+\log (R_o)\right)}{\lambda_a^2},
\]
while, imposing the continuity of $P_{RR}$ at $R=R_i$, we obtain
\[
C_1=\mu_i+\frac{\left(\lambda_a^6-1\right) \mu_o (\log (R_i)-\log (R_o))}{\lambda_a^2}.
\]
In the next Section, we study the stability of the reference configuration with respect to axisymmetric perturbations.

\section{Linear stability analysis}
\label{sec:lin_stab}
In order to characterize the bifurcations exhibited by the elastic body, we exploit the theory of incremental deformations \cite{ogden1997non}. In particular, we introduce a small perturbation of the reference state:
let $\delta\vect{u}$ be the incremental displacement field, we denote by $\boldsymbol{\Gamma}$ its gradient.
We introduce the incremental nominal stress tensor, given by
\begin{equation}
\label{eq:incr_stress}
\begin{aligned}
\delta \tens{P}&=\mathcal{A}:\tens{\Gamma} + p \tens{\Gamma}-\delta p \tens{I},\\
\delta P_{ij} &= A_{ijhl}\Gamma_{lh} + p\,\Gamma_{ij} - \delta p \,\delta_{ij},
\end{aligned}
\end{equation}
where summation over repeated indices is assumed, $\delta p$ is the increment of the Lagrange multiplier $p$, $\delta_{ij}$ is the Kronecker delta, and $\mathcal{A}$ is the fourth-order tensor of elastic moduli. It is defined as
\[
\mathcal{A} = \left.\frac{\partial^2 \Psi}{\partial\tens{F}\partial\tens{F}}\right|_{\tens{F}=\tens{I}} \qquad A_{ijhl} = \left.\frac{\partial^2 \Psi}{\partial F_{ji}\partial F_{lh}}\right|_{
\tens{F}=\tens{I}
}.
\]
 From the expression of the strain energy \eqref{eq:energy}, one obtains $A_{ijhl}=\mu_\beta (F_e)_{i\alpha} (F_e)_{h\alpha} \delta_{jl}$. By linearization of the fully-nonlinear equations \eqref{eq:divP}-\eqref{eq:inc} we get the incremental form of the balance of linear momentum  and of the incompressibility constraint
\begin{equation}
\label{eq:incr_eq}
\left\{
\begin{aligned}
&\Diver\delta\tens{P}=\vect{0},\\
&\tr\tens{\Gamma}=0.
\end{aligned}
\right.
\end{equation}
These partial differential equations are complemented by the following interface and boundary conditions
\begin{equation}
\label{eq:incr_BC}
\left\{
\begin{aligned}
&\delta\tens{P}^T\vect{E}_R=0 &&\text{for }R=R_o,\\
&\llbracket\delta\tens{P}^T\vect{E}_R\rrbracket=0 &&\text{for }R=R_i,\\
&\llbracket\delta\vect{u}\rrbracket=0 &&\text{for }R=R_i.
\end{aligned}
\right.
\end{equation}
Let $\delta\vect{u}$ be an axisymmetric field such that $\delta\vect{u}=u(R,\,Z)\vect{E}_R+w(R,\,Z)\vect{E}_Z$. The following variable separation of the incremental displacement and pressure is assumed:
\begin{equation}
\label{eq:var_sep_UW}
\left\{
\begin{aligned}
&u(R,\,Z)=U(R)\cos(k Z/R_o),\\
&w(R,\,Z)=W(R)\sin(k Z/R_o),\\
&\delta p(R,\,Z)=P(R)\cos(k Z/R_o).
\end{aligned}
\right.
\end{equation}

We first solve analytically the incremental equation in the axoplasm.
Since $A_{ajhl}=\mu_i \delta_{ah} \delta_{jl}$ and the pressure $p$ is constant for $R<R_i$ (see \eqref{eq:pres}), the incremental equations \eqref{eq:incr_eq} reduce to
\begin{widetext}
\[
\left\{
\begin{aligned}
&R R_o \left(k p R W'-R R_o P'+R_o (\mu_i+p) U'+R R_o (\mu_i+p) U''\right)-U \left(k^2 \mu_i R^2+R_o^2 (\mu_i+p)\right)=0,\\
&-k^2 p R W-k^2 \mu_i R W+k R R_o P-k p R R_o U'-k p R_o U+\mu_i R_o^2 W'+\mu_i R R_o^2 W''=0,\\
&k R W+R R_o U'+R_o U=0,
\end{aligned}
\right.
\]
\end{widetext}
where~$'$ denotes the derivative with respect to the radial coordinate.
Following the procedure exposed in \cite{Bigoni_2001}, it is possible to prove that a set of independent solutions, which are continuous at $R=0$ and bounded, is given by
\begin{equation}
\label{eq:sol_core}
\begin{aligned}
&\left\{
\begin{aligned}
&U^1 = I_1\left(\frac{k R}{R_o}\right),\\
&W^1 = -I_0\left(\frac{k R}{R_o}\right),\\
&P^1 = 0,
\end{aligned}
\right.
\\
&\left\{
\begin{aligned}
&U^2 = R I_0\left(\frac{k R}{R_o}\right),\\
&W^2 = -\frac{2 R_o}{k}I_0\left(\frac{k R}{R_o}\right)-R I_1\left(\frac{k R}{R_o}\right),\\
&P^2 = 2\mu_i I_0\left(\frac{k R}{R_o}\right),
\end{aligned}
\right.
\end{aligned}
\end{equation}
where $I_j$ is the modified Bessel function of the first kind of order $j$.

While it is possible to solve analytically the incremental problem in the axoplasm, in the cortex the pressure field $p$ depends on $R$, making the differential equations much more complicated. Nevertheless, they can still be solved numerically.
However, the incremental problem given by \eqref{eq:incr_eq}-\eqref{eq:incr_BC} is numerically stiff and it is convenient to reformulate it in a more suitable form. In particular, we exploit the Stroh formalism \cite{Stroh_1962} to recast the problem into a system of first order differential equations. The Hamiltionian structure of this formulation \cite{Fu_2007} allows us to construct a robust numerical procedure. 
Among the different algorithms that have been proposed in the literature, here we use the impedance matrix method \cite{Norris_2010}, which allows us to write the incremental problem as a differential Riccati equation. Finally, a bifurcation criterion is constructed by enforcing the continuity of the incremental stress and displacement at the interface between the cortex and the axoplasm for non trivial incremental displacements.
The details and the explicit computations are reported in Appendix~\ref{app:stroh}. In the next Section, we show and discuss the outcomes of the stability analysis.

\begin{figure}[b!]
\includegraphics[width=\columnwidth]{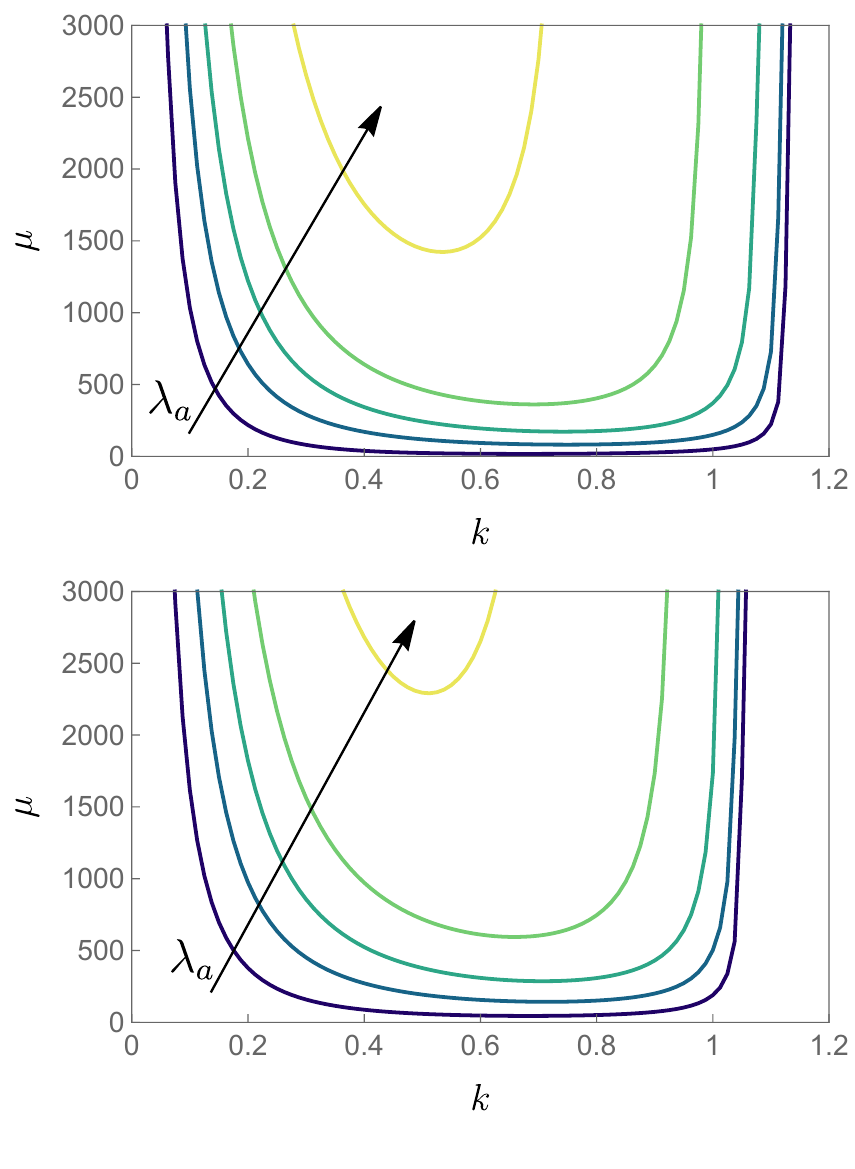}
\caption{Marginal stability curves showing the control parameter $\mu=\mu_o/\mu_i$ versus the dimensionless wave-number $k$ for $\rho=0.8$ (top) and $\rho=0.9$ (bottom), $\lambda_a=0.2,\,0.3,\,0.4,\,0.5,\,0.6$. The arrow denotes the direction in which $\lambda_a$ grows.}
\label{fig:stab_curves}
\end{figure}

\begin{figure}[t!]
\includegraphics[width=\columnwidth]{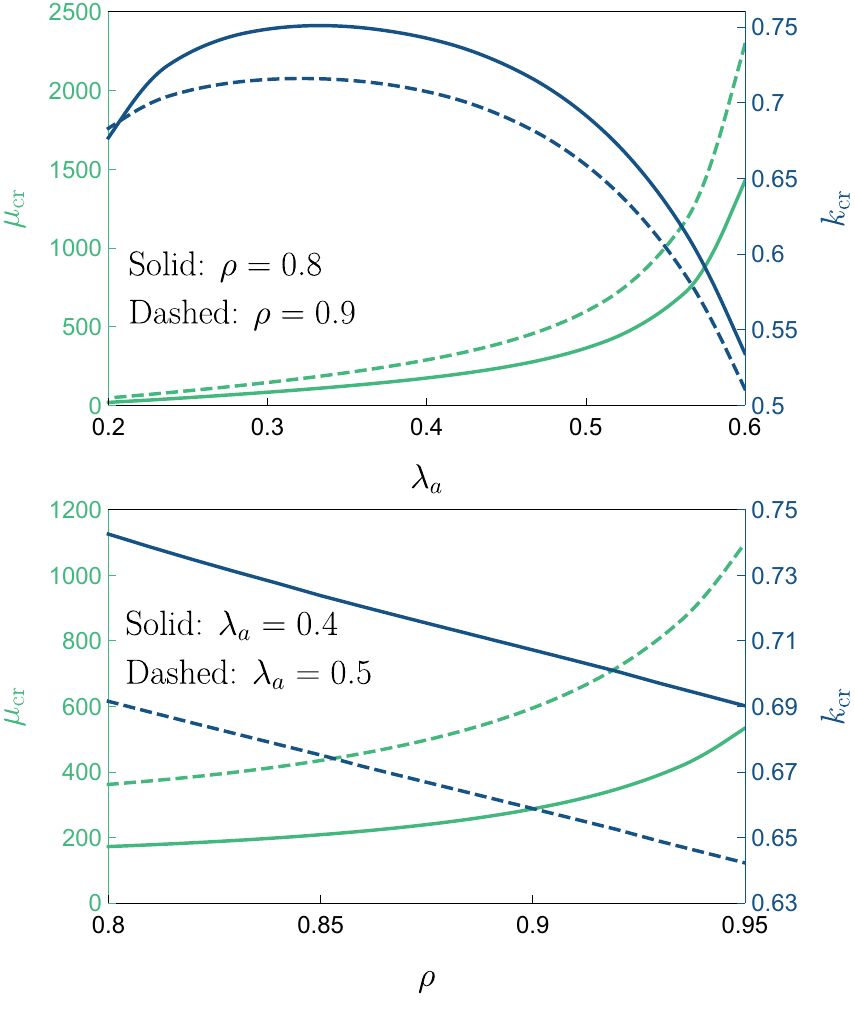}
\caption{Plot of $\mu_\text{cr}$ and $k_\text{cr}$ versus $\lambda_a$ and $\rho$.}
\label{fig:rhol}
\end{figure}

\subsection{Results of the linear stability analysis}

The problem is nondimensionalized with respect to the length scale $R_o$ and the shear modulus $\mu_i$, introducing the aspect ratio $\rho=R_i/R_o$ and the stiffness ratio $\mu=\mu_o/\mu_i$.
When microtubules are depolymerized, the shear modulus of the axoplasm decreases, so that the ratio $\mu$ increases. Thus, it is natural to adopt $\mu$ as control parameter of the bifurcation.

Figure~\ref{fig:stab_curves} shows the marginal stability curves obtained for several values of the active stretch $\lambda_a$. The critical wave-number $k_\text{cr}$ and the critical stiffness ratio $\mu_\text{cr}$ are defined as the coordinates of the minima of the stability curves. Interestingly, in contrast to the elastic Rayleigh-Plateau instability \cite{Mora_2010,Fu_2021}, the linear analysis predicts a finite critical wave-number, see Figure~\ref{fig:rhol}: as the inner shear modulus $\mu_i$ diminishes due to microtubule disruption, the straight axon buckles exhibiting a periodic peristaltic pattern.  The critical wave-number appears to depend linearly on the aspect ratio $\rho$ and, as one could intuitively expect, the critical stiffness ratio diminishes as $\lambda_a$  decreases (i.e. when the actin cortex is more contracted). In general, the critical wave-number belongs to the interval $[0.5,\,0.76]$ for all the considered values of $\rho$ and $\lambda_a$. This means that, depending on $\rho$ and $\lambda_a$, the wave-length of the pattern ranges between $8.37\,R_o$ and $12.56\,R_o$.

Datar and co-workers \cite{Datar_2019} observed that the wave-length of the pearling pattern induced by nocodazole on PC12 neurites increases linearly with the radius of the axon. In particular, the experimentally measured wave-length is $\simeq (11.7761\pm 0.7060)\,R_o$, in agreement with the outcomes of the stability analysis. It is to be remarked that human axon exposed to nocodazole seems to exhibit a longer wave-length \cite{Datar_2019}. This behavior may be caused by the spatially inhomogeneous depolymerization of the cytoskeleton induced by nocodazole: this drug first disrupt the microtubules close to the axonal growth cone, so that only the final part of the axonal shaft exhibits the formation of beads. The study of axonal beading induced by a spatially inhomogeneous depolymerization is beyond the scope of this paper and will be addressed in a future work.

The linear analysis presented in this section can be easily generalized to arbitrary, non axisymmetric perturbations following an analogous procedure. The computations are not reported explicitly but, when the symmetry is broken by the perturbation, the axon appears to be stable.

Compared with previous works on the buckling of layered elastic cylinders, the instability investigated in this paper shows some interesting features. Indeed, the formation of periodic patterning in cylindrical structures induced by active processes, such as growth \cite{Cao_2012,ciarletta2014pattern} or swelling \cite{Dervaux_2011}, has been widely investigated: the surface instability is usually triggered by a coating where the hoop \cite{ciarletta2016morphology} or the axial stress \cite{Du_2019} is compressive. Conversely, in this paper the F-actin cortex contracts in both these directions.

While the linear analysis detects the stability threshold, it does not provide information on the behavior of the buckled axons far away from the bifurcation point. In the next Section, a numerical approximation of the non-linear problem is proposed to overcome this limitation.

\section{Post-buckling analysis}
\label{sec:num}
In order to study the post-buckling evolution of the bifurcated branches, the fully non-linear equations are discretized by means of the finite element method.
The Python library FEniCS is used to implement the numerical code. Assuming axisymmetry, for fixed values of $\lambda_a$ and $\rho$, we use as computational domain the rectangle
\[
\{(X,\,Y)=(Z/R_o,\,R/R_o)\in(O,\,2\pi/k_\text{cr})\times(0,\,1)\},
\]
where $k_\text{cr}$ is the theoretical critical wave-number arising from the linear stability analysis. Periodic boundary conditions are imposed for $X=0$ and $X=2\pi/k_\text{cr}$. Furthermore, the position of the origin is fixed to avoid rigid displacements. 

Using a structured triangular mesh, the displacement and the pressure fields are discretized by using piecewise quadratic polynomials and piecewise constant functions, respectively. Such a mixed formulation is numerically stable for problems arising from incompressible elasticity \cite{boffi2013mixed}. The maximum diameter of the elements is $0.0354$. A small sinusoidal imperfection (having an amplitude of $2.5\cdot 10^{-5}$) is applied to the mesh to trigger the instability.

The code is implemented using the parameter continuation library developed in \cite{Riccobelli_2021}:  starting from $\mu=1$, the control parameter $\mu$ is iteratively incremented of a quantity $\Delta \mu$. The nonlinear problem is solved for using a Newton method, adopting the solution obtained for $\mu$ as initial guess for $\mu+\Delta\mu$.

\begin{figure}[t!]
\includegraphics[width=\columnwidth]{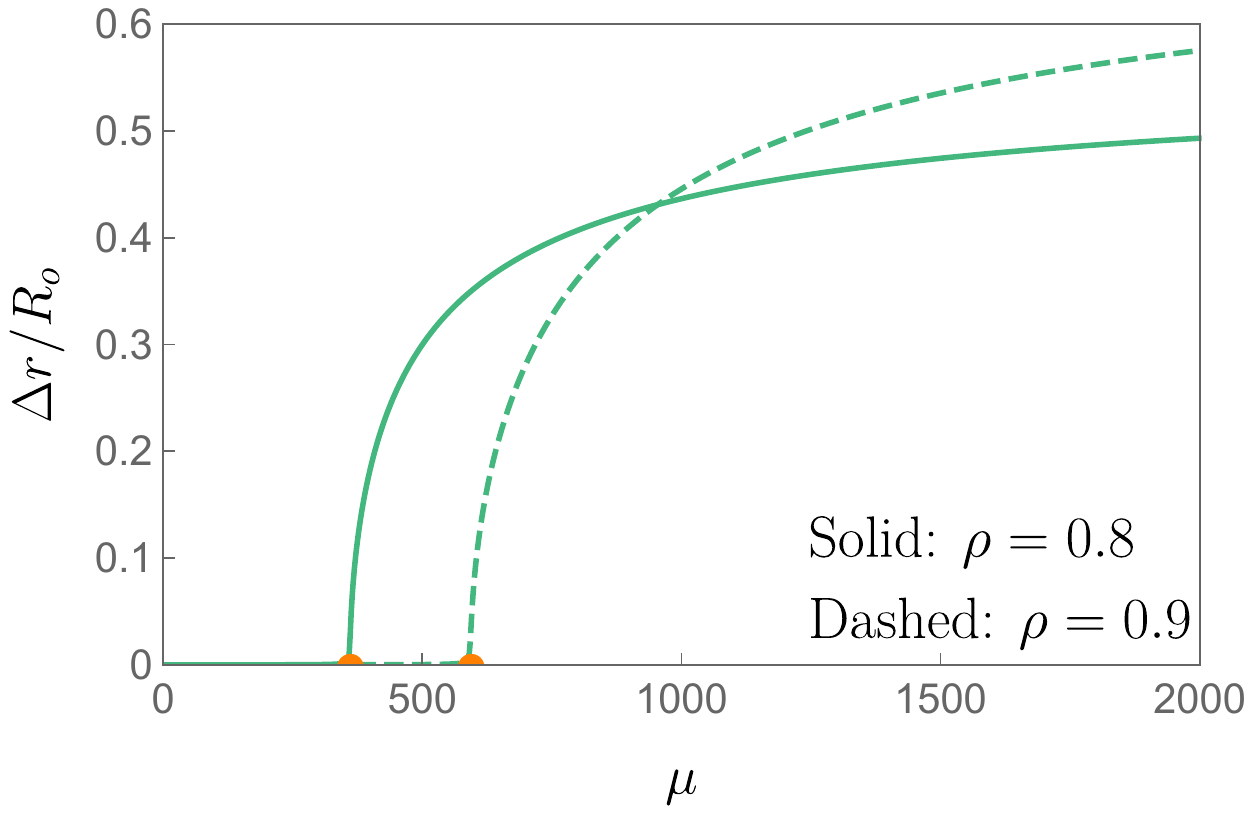}
\caption{Bifurcation diagram showing the normalized beading amplitude $\Delta r/R_o$  versus the control parameter~$\mu$. The solid and dashed lines correspond to two distinct simulations where $\rho=0.8$ and $\rho=0.9$, respectively, while $\lambda_a=0.5$ in both the cases. The orange circles denote the theoretical stability thresholds arising from the linear analysis.}
\label{fig:num}
\end{figure}

\subsection{Results of the numerical simulations}

\begin{figure}[t!]
\includegraphics[width=\columnwidth]{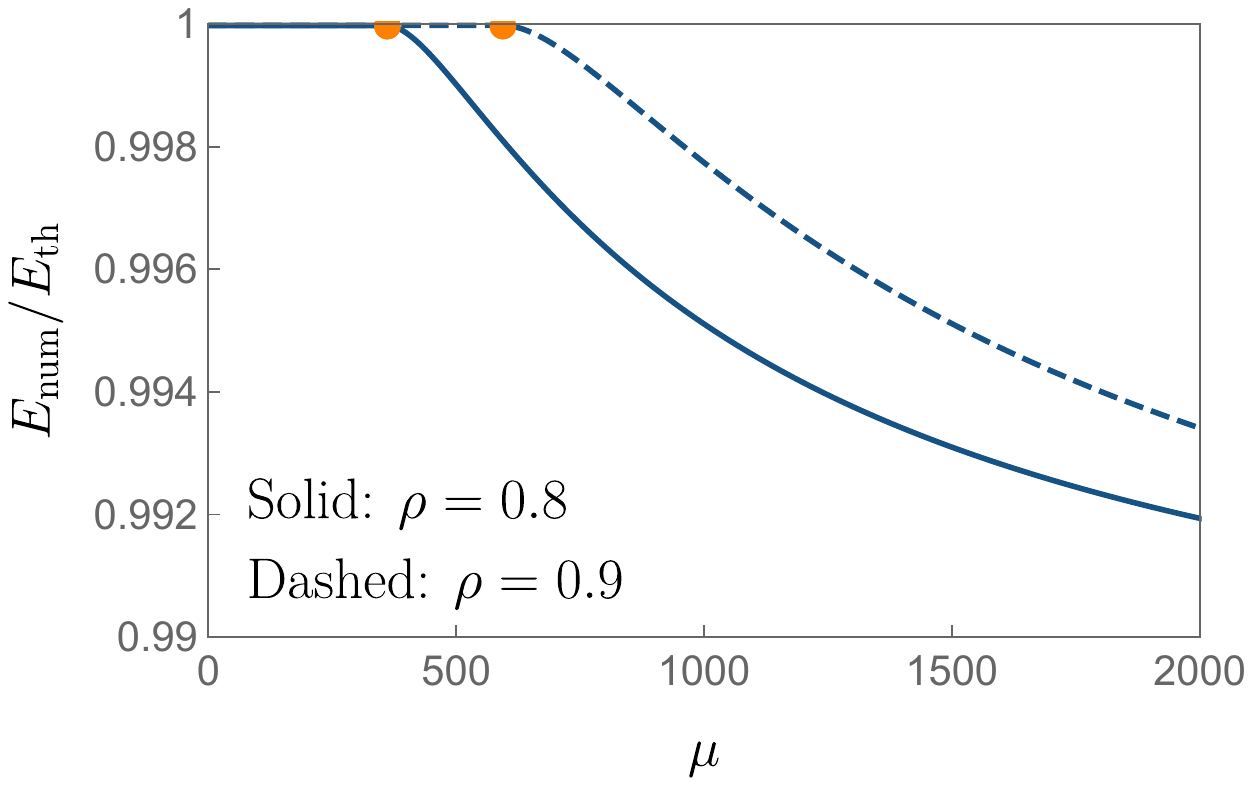}
\caption{Bifurcation diagram showing the energy ratio $E_\text{num}/E_\text{th}$ versus the control parameter~$\mu$. The solid and dashed lines correspond to two distinct simulations where $\rho=0.8$ and $\rho=0.9$, respectively, while $\lambda_a=0.5$ in both the cases. The orange circles denote the theoretical stability thresholds arising from the linear analysis.}
\label{fig:energy}
\end{figure}
\begin{figure}[t!]
\includegraphics[width=\columnwidth]{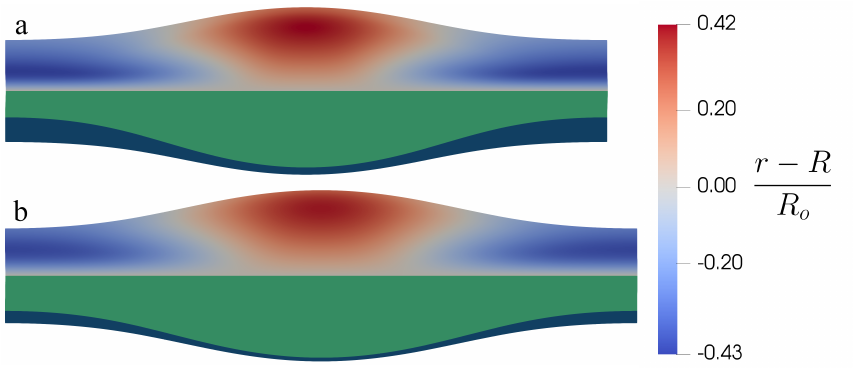}
\caption{Deformed configurations predicted by the finite element simulations for $\rho=0.8$ (snapshot ``a'') and $\rho=0.9$ (snapshot ``b'') when $\mu=2000$, $\lambda_a=0.5$. In the lower part of the axons it is shown the deformed image of the axoplasm (green) and of the actin cortex (blue).}
\label{fig:buckled}
\end{figure}
Let $\Delta r$ be the amplitude of the beading pattern at the free surface, that is
\[
\Delta r = \max_{Z \in [0,\,2 \pi R_o/k_\text{cr}]}r(R_o,\,Z)-\min_{Z \in [0,\,2 \pi R_o/k_\text{cr}]}r(R_o,\,Z).
\]
In Figure~\ref{fig:num}, we depict the bifurcation diagram obtained from the numerical simulations for $\rho=0.8$ and $\rho=0.9$ when the active stretch $\lambda_a$ is $0.5$ \footnote{See Supplemental Material for videos showing the post-buckling evolution predicted by the numerical simulations.}.
We observe that the bifurcation diagrams exhibit the typical shape of a supercritical pitchfork bifurcation, with a continuous increase of $\Delta r/R_o$ at the onset of the instability. We remark that there is a perfect match with the theoretical stability thresholds computed through the linear analysis.
Counterintuitively, despite the marginal stability threshold is higher, the normalized beading amplitude $\Delta r/R_o$ increases faster as the aspect ratio $\rho$ is incremented, resulting in a more pronounced pattern in the nonlinear regime.
Denoting by $E_\text{num}$ and $E_\text{th}$ the energies of the buckled and of the undeformed reference configuration, respectively, in Figure~\ref{fig:energy} we plot the energy ratio $E_\text{num}/E_\text{th}$ versus the control parameter $\mu$.
Finally, the buckled configurations for $\mu=2000$ are reported in Figure~\ref{fig:buckled}. Interestingly, the actin cortex is thinner in correspondence of the bulges provoked by the disruption of the cytoskeleton, while it is thicker where the axonal radius is minimal. 

\section{Concluding remarks}
\label{sec:rem}
\begin{figure*}[t!]
\includegraphics[width=\textwidth]{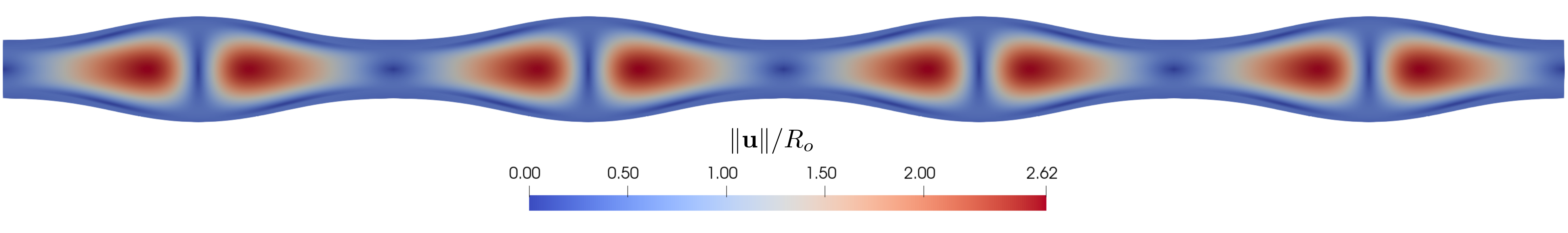}
\caption{Buckled morphology of the axon predicted by the finite element simulations for $\rho=0.9$ when $\mu=2000$, $\lambda_a=0.5$.}
\label{fig:num_long}
\end{figure*}
Summing up, we have characterized the physical mechanisms underlying axonal beading due to microtubule disassembly. Modeling the axonal shaft as a cylindrical bilayer composed of an active hyperelastic material, the reference configuration undergoes a mechanical instability whenever the ratio between the shear modulus of the cortex and of the axoplasm reaches a critical value, that is when the elastic modulus of the axoplasm decreases below a critical threshold. 
The simple model presented in this paper captures the main features of axonal beading: the elasticity of both the F-actin and the microtubules network appears to be fundamental to describe both the onset and the post-buckling evolution of the bifurcated branches. While the wave-length of the instability is controlled by the dimensionless parameters $\lambda_a$ and $\rho$, which are the active strain and the aspect ratio respectively, the amplitude of the pattern is dictated by the ratio between the shear modulus of the cortex and of the axoplasm. The wave-length predicted by the linear analysis is in agreement with experiments performed on PC12 neurites \cite{Datar_2019}.  Furthermore, the postbuckling morphology predicted by the finite element simulations is in qualitative agreement with the experimental one (compare Figure~\ref{fig:sperimentali} (bottom) and Figure~\ref{fig:num_long}).

For the sake of simplicity, in this paper, we have assumed that the depolymerization of the microtubules is spatially homogeneous. However, in some cases the disruption of the cytoskeleton is faster close to the growth cone, as happens in human axons exposed to nocodazole \cite{Datar_2019}. This can lead to a spatial modulation of the wave-length of the pattern that will be studied in a future work. Furthermore, future efforts will be devoted to study the influence of the axoplasm poroelasticity on the shape transitions exhibited by the axon. In fact, the axoplasm is composed of both a solid and a fluid phase. Modeling the axon as a poroelastic mixture, where the microtubule depolymerization gives rise to an exchange of mass between the liquid and solid phase, may lead to a better fit of the experimental shapes even when the depolymerization is spatially inhomogeneous. Another interesting aspect that deserves further study is the role of F-actin disruption in the process of axonal retraction \cite{Datar_2019}.

\appendix
\section{Stroh formulation and impedance matrix method}
\label{app:stroh}
In this appendix, we first rewrite the incremental equations \eqref{eq:incr_eq} as a system of first order differential equations exhibiting an Hamiltionian structure using the Stroh formalism  \cite{Stroh_1962}.
This technique requires to consider $\delta P_{RR}$ and $\delta P_{RZ}$ as additional unknowns of the problem, assuming the variable separation
\begin{align}
\label{eq:var_sep_stress1}
&\delta P_{RR}(R,\,Z)=S_{RR}(R)\cos(k Z/R_o),\\
\label{eq:var_sep_stress2}
&\delta P_{RZ}(R,\,Z)=S_{RZ}(R)\sin(k Z/R_o).
\end{align}
By substituting \eqref{eq:var_sep_stress1} into \eqref{eq:incr_stress}, we obtain an expression for the pressure $P(R)$ (see \eqref{eq:var_sep_UW})
\[
P(R) = U'(R) \left(\lambda_a^4 \mu_o+p(R)\right)-S_{RR}(R).
\]

It remains to determine a system of four equations for the unknowns $U,\,V,\,S_{RR},\,S_{RZ}$. These equations are the incremental form of the incompressibility constraint and of the balance of the linear momentum (three scalar equations), see \eqref{eq:incr_eq}, and the constitutive equation for $\delta P_{RZ}$, given by \eqref{eq:incr_stress}. They can be written as the following system of ordinary differential equations:
\begin{equation}
\label{eq:Stroh}
\boldsymbol{\eta}'=\frac{1}{R}\tens{N}\boldsymbol{\eta},
\end{equation}
where $\boldsymbol{\eta}=[U,\,W,\,R S_{RR},\,R S_{RZ}]$ and $\tens{N}(R)$ is the $4\times4$ Stroh matrix, having the following block form
\begin{equation}
\label{eq:Stroh_block}
\tens{N}=\begin{bmatrix}
\tens{N}_1 &\tens{N}_2\\
\tens{N}_3 &-\tens{N}_1^T
\end{bmatrix},
\end{equation}
where $\tens{N}_1,\,\tens{N}_2,\,\tens{N}_3$ are $2\times2$ matrices such that $\tens{N}_2=\tens{N}_2^T$ and $\tens{N}_3=\tens{N}_3^T$. Their expressions are given by
\begin{gather*}
\tens{N}_1=
\begin{bmatrix}
 -1 & -\dfrac{k R}{R_o} \\
 \dfrac{k R p}{\lambda_a^4 \mu_o R_o} & 0 \\
\end{bmatrix},\qquad
\tens{N}_2=
\begin{bmatrix}
 0 & 0 \\
 0 & \dfrac{1}{\lambda_a^4 \mu_o}
\end{bmatrix},\\
\tens{N}_3=
\begin{bmatrix}
\alpha_1 &  \alpha_2\\
 \alpha_2 & \alpha_3 \\
\end{bmatrix},
\end{gather*}
where
\begin{gather*}
\alpha_1= p \left(2-\dfrac{k^2 R^2 p}{\lambda_a^4 \mu_o R_o^2}\right)+\dfrac{\mu_o \left(\dfrac{k^2 R^2}{R_o^2}+\lambda_a^6+1\right)}{\lambda_a^2},\\
\alpha_2=\dfrac{k R \left(\lambda_a^4 \mu_o+p\right)}{R_o},\\
\alpha_3=\dfrac{k^2 R^2 \left(\lambda_a^6 \mu_o+\mu_o+2 \lambda_a^2 p\right)}{\lambda_a^2 R_o^2}.
\end{gather*}

We can now numerically solve the incremental problem by using the equation \eqref{eq:Stroh}. A very robust numerical scheme is based on the impedance matrix method. More explicitly, we introduce the conditional impedance matrix $\tens{Z}(R,\,R_o)$ \cite{Norris_2010}, so that 
\begin{equation}
\label{eq:cond_mat}
R\vect{S}(R) = \tens{Z}(R,\,R_o)\vect{U}(R),
\end{equation}
where $\vect{U}=[U,\,W]$ and $\vect{S}=[S_{RR},\,S_{RZ}]$.
Plugging \eqref{eq:cond_mat} in \eqref{eq:Stroh}, we get the following equations
\begin{align}
\label{eqn:U'}
\vect{U}'&=\frac{1}{R}(\tens{N}_1\vect{U}+\tens{N}_2\tens{Z}\vect{U}),\\
\label{eqn:ZU'}
\tens{Z}'\vect{U}+\tens{Z}\vect{U}'&=\frac{1}{R}(\tens{N}_3\vect{U}-\tens{N}_1^T\tens{Z}\vect{U}).
\end{align}
Substituting \eqref{eqn:U'} into \eqref{eqn:ZU'}, we obtain a Riccati differential equation:
\begin{equation}
\label{eq:Riccati}
R\tens{Z}'=-\tens{Z}\tens{N}_1-\tens{N}_1^T\tens{Z}-\tens{Z}\tens{N}_2\tens{Z}+\tens{N}_3.
\end{equation}
The Riccati equation is complemented by the the initial condition $\tens{Z}(R_o,\,R_o)=\tens{0}$, corresponding to the boundary condition $\delta\tens{P}\vect{E}_R=\vect{0}$ for $R=R_o$ \cite{Norris_2010}.

Finally, a bifurcation criterion is constructed by enforcing the continuity of the incremental stress and displacement at the interface. Identifying with
\[
\begin{aligned}
&\vect{U}_i=\lim_{R\rightarrow R_i^-}\vect{U}(R),
&&\vect{S}_i=\lim_{R\rightarrow R_i^-}\vect{S}(R),\\
&\vect{U}_o=\lim_{R\rightarrow R_i^+}\vect{U}(R),
&&\vect{S}_o=\lim_{R\rightarrow R_i^+}\vect{S}(R),
\end{aligned}
\]
then, from \eqref{eq:incr_BC} and \eqref{eq:cond_mat}, we obtain
\begin{equation}
\label{eq:cisiamoquasi}
\vect{S}_i = \vect{S}_o = \tens{Z}(R_i,\,R_o)\vect{U}_o = \tens{Z}(R_i,\,R_o)\vect{U}_i.
\end{equation}
Using \eqref{eq:incr_stress}, \eqref{eq:var_sep_stress1} and \eqref{eq:var_sep_stress2}, denoting by $S_{RR}^1,\,S_{RR}^2,\,S_{RZ}^1,\,S_{RZ}^2$ the components of $\vect{S}$ corresponding to the solutions \eqref{eq:sol_core} for the axoplasm, we introduce the matrices $\tens{\Sigma}(R)$ and $\tens{\Omega}(R)$, defined as
\[
\begin{gathered}
\tens{\Sigma}(R)=\begin{bmatrix}
S_{RR}^1(R) &S_{RR}^2(R)\\
S_{RZ}^1(R) &S_{RZ}^2(R)\\
\end{bmatrix},\\
\tens{\Omega}(R)=\begin{bmatrix}
U^1(R) &U^2(R)\\
W^1(R) &W^2(R)\\
\end{bmatrix}.
\end{gathered}
\]

From \eqref{eq:cisiamoquasi}, non-trivial solutions exist whenever \cite{Balbi_2020, Riccobelli_2020}
\begin{equation}
\label{eq:stopcond}
\det\bigg(\tens{\Sigma}(R_i)-\tens{Z}(R_i,\,R_o)\tens{\Omega}(R_i)\bigg)=0.
\end{equation} 
The Riccati equation \eqref{eq:Riccati} is integrated numerically from $R_o$ to $R_i$ using the software \textsc{Mathematica} (version 12.2), incrementing the control parameter $\mu$ for fixed values of $\lambda_a$, $k$ and $\rho$ until the bifurcation criterion \eqref{eq:stopcond} is reached. 

\begin{acknowledgments}
I thank Dr.~G.~Bevilacqua, Prof.~P.~Ciarletta, Prof.~G.~Noselli, and Prof.~P.~A.~Pullarkat for fruitful discussions.
This work has been supported by National
Group of Mathematical Physics (GNFM -- INdAM) through
the program \emph{Progetto Giovani 2020} and  by Regione Lombardia project NEWMED (Grant No. POR FESR 2014-2020).
\end{acknowledgments}

%


\begin{thebibliography}{47}%
\makeatletter
\providecommand \@ifxundefined [1]{%
 \@ifx{#1\undefined}
}%
\providecommand \@ifnum [1]{%
 \ifnum #1\expandafter \@firstoftwo
 \else \expandafter \@secondoftwo
 \fi
}%
\providecommand \@ifx [1]{%
 \ifx #1\expandafter \@firstoftwo
 \else \expandafter \@secondoftwo
 \fi
}%
\providecommand \natexlab [1]{#1}%
\providecommand \enquote  [1]{``#1''}%
\providecommand \bibnamefont  [1]{#1}%
\providecommand \bibfnamefont [1]{#1}%
\providecommand \citenamefont [1]{#1}%
\providecommand \href@noop [0]{\@secondoftwo}%
\providecommand \href [0]{\begingroup \@sanitize@url \@href}%
\providecommand \@href[1]{\@@startlink{#1}\@@href}%
\providecommand \@@href[1]{\endgroup#1\@@endlink}%
\providecommand \@sanitize@url [0]{\catcode `\\12\catcode `\$12\catcode
  `\&12\catcode `\#12\catcode `\^12\catcode `\_12\catcode `\%12\relax}%
\providecommand \@@startlink[1]{}%
\providecommand \@@endlink[0]{}%
\providecommand \url  [0]{\begingroup\@sanitize@url \@url }%
\providecommand \@url [1]{\endgroup\@href {#1}{\urlprefix }}%
\providecommand \urlprefix  [0]{URL }%
\providecommand \Eprint [0]{\href }%
\providecommand \doibase [0]{https://doi.org/}%
\providecommand \selectlanguage [0]{\@gobble}%
\providecommand \bibinfo  [0]{\@secondoftwo}%
\providecommand \bibfield  [0]{\@secondoftwo}%
\providecommand \translation [1]{[#1]}%
\providecommand \BibitemOpen [0]{}%
\providecommand \bibitemStop [0]{}%
\providecommand \bibitemNoStop [0]{.\EOS\space}%
\providecommand \EOS [0]{\spacefactor3000\relax}%
\providecommand \BibitemShut  [1]{\csname bibitem#1\endcsname}%
\let\auto@bib@innerbib\@empty
\bibitem [{\citenamefont {Beach}\ \emph {et~al.}(2020)\citenamefont {Beach},
  \citenamefont {Praschan}, \citenamefont {Hogan}, \citenamefont {Dotson},
  \citenamefont {Merideth}, \citenamefont {Kontos}, \citenamefont
  {Fricchione},\ and\ \citenamefont {Smith}}]{Beach_2020}%
  \BibitemOpen
  \bibfield  {author} {\bibinfo {author} {\bibfnamefont {S.~R.}\ \bibnamefont
  {Beach}}, \bibinfo {author} {\bibfnamefont {N.~C.}\ \bibnamefont {Praschan}},
  \bibinfo {author} {\bibfnamefont {C.}~\bibnamefont {Hogan}}, \bibinfo
  {author} {\bibfnamefont {S.}~\bibnamefont {Dotson}}, \bibinfo {author}
  {\bibfnamefont {F.}~\bibnamefont {Merideth}}, \bibinfo {author}
  {\bibfnamefont {N.}~\bibnamefont {Kontos}}, \bibinfo {author} {\bibfnamefont
  {G.~L.}\ \bibnamefont {Fricchione}},\ and\ \bibinfo {author} {\bibfnamefont
  {F.~A.}\ \bibnamefont {Smith}},\ }\bibfield  {title} {\bibinfo {title}
  {Delirium in {COVID}-19: A case series and exploration of potential
  mechanisms for central nervous system involvement},\ }\href
  {https://doi.org/10.1016/j.genhosppsych.2020.05.008} {\bibfield  {journal}
  {\bibinfo  {journal} {Gen. Hosp. Psychiatry}\ }\textbf {\bibinfo {volume}
  {65}},\ \bibinfo {pages} {47} (\bibinfo {year} {2020})}\BibitemShut {NoStop}%
\bibitem [{\citenamefont {Troyer}\ \emph {et~al.}(2020)\citenamefont {Troyer},
  \citenamefont {Kohn},\ and\ \citenamefont {Hong}}]{Troyer_2020}%
  \BibitemOpen
  \bibfield  {author} {\bibinfo {author} {\bibfnamefont {E.~A.}\ \bibnamefont
  {Troyer}}, \bibinfo {author} {\bibfnamefont {J.~N.}\ \bibnamefont {Kohn}},\
  and\ \bibinfo {author} {\bibfnamefont {S.}~\bibnamefont {Hong}},\ }\bibfield
  {title} {\bibinfo {title} {Are we facing a crashing wave of neuropsychiatric
  sequelae of {COVID}-19? neuropsychiatric symptoms and potential immunologic
  mechanisms},\ }\href {https://doi.org/10.1016/j.bbi.2020.04.027} {\bibfield
  {journal} {\bibinfo  {journal} {Brain Behav. Immun.}\ }\textbf {\bibinfo
  {volume} {87}},\ \bibinfo {pages} {34} (\bibinfo {year} {2020})}\BibitemShut
  {NoStop}%
\bibitem [{\citenamefont {Ellul}\ \emph {et~al.}(2020)\citenamefont {Ellul},
  \citenamefont {Benjamin}, \citenamefont {Singh}, \citenamefont {Lant},
  \citenamefont {Michael}, \citenamefont {Easton}, \citenamefont {Kneen},
  \citenamefont {Defres}, \citenamefont {Sejvar},\ and\ \citenamefont
  {Solomon}}]{ELLUL2020767}%
  \BibitemOpen
  \bibfield  {author} {\bibinfo {author} {\bibfnamefont {M.~A.}\ \bibnamefont
  {Ellul}}, \bibinfo {author} {\bibfnamefont {L.}~\bibnamefont {Benjamin}},
  \bibinfo {author} {\bibfnamefont {B.}~\bibnamefont {Singh}}, \bibinfo
  {author} {\bibfnamefont {S.}~\bibnamefont {Lant}}, \bibinfo {author}
  {\bibfnamefont {B.~D.}\ \bibnamefont {Michael}}, \bibinfo {author}
  {\bibfnamefont {A.}~\bibnamefont {Easton}}, \bibinfo {author} {\bibfnamefont
  {R.}~\bibnamefont {Kneen}}, \bibinfo {author} {\bibfnamefont
  {S.}~\bibnamefont {Defres}}, \bibinfo {author} {\bibfnamefont
  {J.}~\bibnamefont {Sejvar}},\ and\ \bibinfo {author} {\bibfnamefont
  {T.}~\bibnamefont {Solomon}},\ }\bibfield  {title} {\bibinfo {title}
  {Neurological associations of covid-19},\ }\href
  {https://doi.org/https://doi.org/10.1016/S1474-4422(20)30221-0} {\bibfield
  {journal} {\bibinfo  {journal} {Lancet Neurol.}\ }\textbf {\bibinfo {volume}
  {19}},\ \bibinfo {pages} {767} (\bibinfo {year} {2020})}\BibitemShut
  {NoStop}%
\bibitem [{\citenamefont {Rogers}\ \emph {et~al.}(2020)\citenamefont {Rogers},
  \citenamefont {Chesney}, \citenamefont {Oliver}, \citenamefont {Pollak},
  \citenamefont {McGuire}, \citenamefont {Fusar-Poli}, \citenamefont {Zandi},
  \citenamefont {Lewis},\ and\ \citenamefont {David}}]{Rogers_2020}%
  \BibitemOpen
  \bibfield  {author} {\bibinfo {author} {\bibfnamefont {J.~P.}\ \bibnamefont
  {Rogers}}, \bibinfo {author} {\bibfnamefont {E.}~\bibnamefont {Chesney}},
  \bibinfo {author} {\bibfnamefont {D.}~\bibnamefont {Oliver}}, \bibinfo
  {author} {\bibfnamefont {T.~A.}\ \bibnamefont {Pollak}}, \bibinfo {author}
  {\bibfnamefont {P.}~\bibnamefont {McGuire}}, \bibinfo {author} {\bibfnamefont
  {P.}~\bibnamefont {Fusar-Poli}}, \bibinfo {author} {\bibfnamefont {M.~S.}\
  \bibnamefont {Zandi}}, \bibinfo {author} {\bibfnamefont {G.}~\bibnamefont
  {Lewis}},\ and\ \bibinfo {author} {\bibfnamefont {A.~S.}\ \bibnamefont
  {David}},\ }\bibfield  {title} {\bibinfo {title} {Psychiatric and
  neuropsychiatric presentations associated with severe coronavirus infections:
  a systematic review and meta-analysis with comparison to the {COVID}-19
  pandemic},\ }\href {https://doi.org/10.1016/s2215-0366(20)30203-0} {\bibfield
   {journal} {\bibinfo  {journal} {Lancet Psychiatry}\ }\textbf {\bibinfo
  {volume} {7}},\ \bibinfo {pages} {611} (\bibinfo {year} {2020})}\BibitemShut
  {NoStop}%
\bibitem [{\citenamefont {Jacomy}\ \emph {et~al.}(2006)\citenamefont {Jacomy},
  \citenamefont {Fragoso}, \citenamefont {Almazan}, \citenamefont {Mushynski},\
  and\ \citenamefont {Talbot}}]{Jacomy_2006}%
  \BibitemOpen
  \bibfield  {author} {\bibinfo {author} {\bibfnamefont {H.}~\bibnamefont
  {Jacomy}}, \bibinfo {author} {\bibfnamefont {G.}~\bibnamefont {Fragoso}},
  \bibinfo {author} {\bibfnamefont {G.}~\bibnamefont {Almazan}}, \bibinfo
  {author} {\bibfnamefont {W.~E.}\ \bibnamefont {Mushynski}},\ and\ \bibinfo
  {author} {\bibfnamefont {P.~J.}\ \bibnamefont {Talbot}},\ }\bibfield  {title}
  {\bibinfo {title} {Human coronavirus {OC}43 infection induces chronic
  encephalitis leading to disabilities in {BALB}/c mice},\ }\href
  {https://doi.org/10.1016/j.virol.2006.01.049} {\bibfield  {journal} {\bibinfo
   {journal} {Virology}\ }\textbf {\bibinfo {volume} {349}},\ \bibinfo {pages}
  {335} (\bibinfo {year} {2006})}\BibitemShut {NoStop}%
\bibitem [{\citenamefont {Niki{\'c}}\ \emph {et~al.}(2011)\citenamefont
  {Niki{\'c}}, \citenamefont {Merkler}, \citenamefont {Sorbara}, \citenamefont
  {Brinkoetter}, \citenamefont {Kreutzfeldt}, \citenamefont {Bareyre},
  \citenamefont {Br{\"u}ck}, \citenamefont {Bishop}, \citenamefont {Misgeld},\
  and\ \citenamefont {Kerschensteiner}}]{nikic2011reversible}%
  \BibitemOpen
  \bibfield  {author} {\bibinfo {author} {\bibfnamefont {I.}~\bibnamefont
  {Niki{\'c}}}, \bibinfo {author} {\bibfnamefont {D.}~\bibnamefont {Merkler}},
  \bibinfo {author} {\bibfnamefont {C.}~\bibnamefont {Sorbara}}, \bibinfo
  {author} {\bibfnamefont {M.}~\bibnamefont {Brinkoetter}}, \bibinfo {author}
  {\bibfnamefont {M.}~\bibnamefont {Kreutzfeldt}}, \bibinfo {author}
  {\bibfnamefont {F.~M.}\ \bibnamefont {Bareyre}}, \bibinfo {author}
  {\bibfnamefont {W.}~\bibnamefont {Br{\"u}ck}}, \bibinfo {author}
  {\bibfnamefont {D.}~\bibnamefont {Bishop}}, \bibinfo {author} {\bibfnamefont
  {T.}~\bibnamefont {Misgeld}},\ and\ \bibinfo {author} {\bibfnamefont
  {M.}~\bibnamefont {Kerschensteiner}},\ }\bibfield  {title} {\bibinfo {title}
  {A reversible form of axon damage in experimental autoimmune
  encephalomyelitis and multiple sclerosis},\ }\href@noop {} {\bibfield
  {journal} {\bibinfo  {journal} {Nat. Med.}\ }\textbf {\bibinfo {volume}
  {17}},\ \bibinfo {pages} {495} (\bibinfo {year} {2011})}\BibitemShut
  {NoStop}%
\bibitem [{\citenamefont {Stokin}\ \emph {et~al.}(2005)\citenamefont {Stokin},
  \citenamefont {Lillo}, \citenamefont {Falzone}, \citenamefont {Brusch},
  \citenamefont {Rockenstein}, \citenamefont {Mount}, \citenamefont {Raman},
  \citenamefont {Davies}, \citenamefont {Masliah}, \citenamefont {Williams},\
  and\ \citenamefont {Goldstein}}]{Stokin1282}%
  \BibitemOpen
  \bibfield  {author} {\bibinfo {author} {\bibfnamefont {G.~B.}\ \bibnamefont
  {Stokin}}, \bibinfo {author} {\bibfnamefont {C.}~\bibnamefont {Lillo}},
  \bibinfo {author} {\bibfnamefont {T.~L.}\ \bibnamefont {Falzone}}, \bibinfo
  {author} {\bibfnamefont {R.~G.}\ \bibnamefont {Brusch}}, \bibinfo {author}
  {\bibfnamefont {E.}~\bibnamefont {Rockenstein}}, \bibinfo {author}
  {\bibfnamefont {S.~L.}\ \bibnamefont {Mount}}, \bibinfo {author}
  {\bibfnamefont {R.}~\bibnamefont {Raman}}, \bibinfo {author} {\bibfnamefont
  {P.}~\bibnamefont {Davies}}, \bibinfo {author} {\bibfnamefont
  {E.}~\bibnamefont {Masliah}}, \bibinfo {author} {\bibfnamefont {D.~S.}\
  \bibnamefont {Williams}},\ and\ \bibinfo {author} {\bibfnamefont {L.~S.~B.}\
  \bibnamefont {Goldstein}},\ }\bibfield  {title} {\bibinfo {title} {Axonopathy
  and transport deficits early in the pathogenesis of
  alzheimer{\textquoteright}s disease},\ }\href
  {https://doi.org/10.1126/science.1105681} {\bibfield  {journal} {\bibinfo
  {journal} {Science}\ }\textbf {\bibinfo {volume} {307}},\ \bibinfo {pages}
  {1282} (\bibinfo {year} {2005})}\BibitemShut {NoStop}%
\bibitem [{\citenamefont {Tagliaferro}\ and\ \citenamefont
  {Burke}(2016)}]{tagliaferro2016retrograde}%
  \BibitemOpen
  \bibfield  {author} {\bibinfo {author} {\bibfnamefont {P.}~\bibnamefont
  {Tagliaferro}}\ and\ \bibinfo {author} {\bibfnamefont {R.~E.}\ \bibnamefont
  {Burke}},\ }\bibfield  {title} {\bibinfo {title} {Retrograde axonal
  degeneration in parkinson disease},\ }\href@noop {} {\bibfield  {journal}
  {\bibinfo  {journal} {J. Parkinsons Dis.}\ }\textbf {\bibinfo {volume} {6}},\
  \bibinfo {pages} {1} (\bibinfo {year} {2016})}\BibitemShut {NoStop}%
\bibitem [{\citenamefont {Bain}\ and\ \citenamefont
  {Meaney}(2000)}]{bain2000tissue}%
  \BibitemOpen
  \bibfield  {author} {\bibinfo {author} {\bibfnamefont {A.~C.}\ \bibnamefont
  {Bain}}\ and\ \bibinfo {author} {\bibfnamefont {D.~F.}\ \bibnamefont
  {Meaney}},\ }\bibfield  {title} {\bibinfo {title} {Tissue-level thresholds
  for axonal damage in an experimental model of central nervous system white
  matter injury},\ }\href@noop {} {\bibfield  {journal} {\bibinfo  {journal}
  {J. Biomech. Eng.}\ }\textbf {\bibinfo {volume} {122}},\ \bibinfo {pages}
  {615} (\bibinfo {year} {2000})}\BibitemShut {NoStop}%
\bibitem [{\citenamefont {Kolaric}\ \emph {et~al.}(2013)\citenamefont
  {Kolaric}, \citenamefont {Thomson}, \citenamefont {Edgar},\ and\
  \citenamefont {Brown}}]{Kolaric_2013}%
  \BibitemOpen
  \bibfield  {author} {\bibinfo {author} {\bibfnamefont {K.~V.}\ \bibnamefont
  {Kolaric}}, \bibinfo {author} {\bibfnamefont {G.}~\bibnamefont {Thomson}},
  \bibinfo {author} {\bibfnamefont {J.~M.}\ \bibnamefont {Edgar}},\ and\
  \bibinfo {author} {\bibfnamefont {A.~M.}\ \bibnamefont {Brown}},\ }\bibfield
  {title} {\bibinfo {title} {Focal axonal swellings and associated
  ultrastructural changes attenuate conduction velocity in central nervous
  system axons: a computer modeling study},\ }\bibfield  {journal} {\bibinfo
  {journal} {Physiol. Rep.}\ }\textbf {\bibinfo {volume} {1}},\ \href
  {https://doi.org/10.1002/phy2.59} {10.1002/phy2.59} (\bibinfo {year}
  {2013})\BibitemShut {NoStop}%
\bibitem [{\citenamefont {Nunomura}\ \emph {et~al.}(2006)\citenamefont
  {Nunomura}, \citenamefont {Castellani}, \citenamefont {Zhu}, \citenamefont
  {Moreira}, \citenamefont {Perry},\ and\ \citenamefont
  {Smith}}]{nunomura2006involvement}%
  \BibitemOpen
  \bibfield  {author} {\bibinfo {author} {\bibfnamefont {A.}~\bibnamefont
  {Nunomura}}, \bibinfo {author} {\bibfnamefont {R.~J.}\ \bibnamefont
  {Castellani}}, \bibinfo {author} {\bibfnamefont {X.}~\bibnamefont {Zhu}},
  \bibinfo {author} {\bibfnamefont {P.~I.}\ \bibnamefont {Moreira}}, \bibinfo
  {author} {\bibfnamefont {G.}~\bibnamefont {Perry}},\ and\ \bibinfo {author}
  {\bibfnamefont {M.~A.}\ \bibnamefont {Smith}},\ }\bibfield  {title} {\bibinfo
  {title} {Involvement of oxidative stress in alzheimer disease},\ }\href@noop
  {} {\bibfield  {journal} {\bibinfo  {journal} {J. Neuropathol. Exp.}\
  }\textbf {\bibinfo {volume} {65}},\ \bibinfo {pages} {631} (\bibinfo {year}
  {2006})}\BibitemShut {NoStop}%
\bibitem [{\citenamefont {Roediger}\ and\ \citenamefont
  {Armati}(2003)}]{Roediger_2003}%
  \BibitemOpen
  \bibfield  {author} {\bibinfo {author} {\bibfnamefont {B.}~\bibnamefont
  {Roediger}}\ and\ \bibinfo {author} {\bibfnamefont {P.~J.}\ \bibnamefont
  {Armati}},\ }\bibfield  {title} {\bibinfo {title} {Oxidative stress induces
  axonal beading in cultured human brain tissue},\ }\href
  {https://doi.org/10.1016/s0969-9961(03)00038-x} {\bibfield  {journal}
  {\bibinfo  {journal} {Neurobiol. Dis.}\ }\textbf {\bibinfo {volume} {13}},\
  \bibinfo {pages} {222} (\bibinfo {year} {2003})}\BibitemShut {NoStop}%
\bibitem [{\citenamefont {Woerman}\ \emph {et~al.}(2016)\citenamefont
  {Woerman}, \citenamefont {Aoyagi}, \citenamefont {Patel}, \citenamefont
  {Kazmi}, \citenamefont {Lobach}, \citenamefont {Grinberg}, \citenamefont
  {McKee}, \citenamefont {Seeley}, \citenamefont {Olson},\ and\ \citenamefont
  {Prusiner}}]{Woerman_2016}%
  \BibitemOpen
  \bibfield  {author} {\bibinfo {author} {\bibfnamefont {A.~L.}\ \bibnamefont
  {Woerman}}, \bibinfo {author} {\bibfnamefont {A.}~\bibnamefont {Aoyagi}},
  \bibinfo {author} {\bibfnamefont {S.}~\bibnamefont {Patel}}, \bibinfo
  {author} {\bibfnamefont {S.~A.}\ \bibnamefont {Kazmi}}, \bibinfo {author}
  {\bibfnamefont {I.}~\bibnamefont {Lobach}}, \bibinfo {author} {\bibfnamefont
  {L.~T.}\ \bibnamefont {Grinberg}}, \bibinfo {author} {\bibfnamefont {A.~C.}\
  \bibnamefont {McKee}}, \bibinfo {author} {\bibfnamefont {W.~W.}\ \bibnamefont
  {Seeley}}, \bibinfo {author} {\bibfnamefont {S.~H.}\ \bibnamefont {Olson}},\
  and\ \bibinfo {author} {\bibfnamefont {S.~B.}\ \bibnamefont {Prusiner}},\
  }\bibfield  {title} {\bibinfo {title} {Tau prions from alzheimer's disease
  and chronic traumatic encephalopathy patients propagate in cultured cells},\
  }\href {https://doi.org/10.1073/pnas.1616344113} {\bibfield  {journal}
  {\bibinfo  {journal} {Proceedings of the National Academy of Sciences}\
  }\textbf {\bibinfo {volume} {113}},\ \bibinfo {pages} {E8187} (\bibinfo
  {year} {2016})}\BibitemShut {NoStop}%
\bibitem [{\citenamefont {van~den Bedem}\ and\ \citenamefont
  {Kuhl}(2015)}]{van_den_Bedem_2015}%
  \BibitemOpen
  \bibfield  {author} {\bibinfo {author} {\bibfnamefont {H.}~\bibnamefont
  {van~den Bedem}}\ and\ \bibinfo {author} {\bibfnamefont {E.}~\bibnamefont
  {Kuhl}},\ }\bibfield  {title} {\bibinfo {title} {Tau-ism: The yin and yang of
  microtubule sliding, detachment, and rupture},\ }\href
  {https://doi.org/10.1016/j.bpj.2015.10.020} {\bibfield  {journal} {\bibinfo
  {journal} {Biophys. J.}\ }\textbf {\bibinfo {volume} {109}},\ \bibinfo
  {pages} {2215} (\bibinfo {year} {2015})}\BibitemShut {NoStop}%
\bibitem [{\citenamefont {de~Rooij}\ and\ \citenamefont
  {Kuhl}(2018)}]{de_Rooij_2018}%
  \BibitemOpen
  \bibfield  {author} {\bibinfo {author} {\bibfnamefont {R.}~\bibnamefont
  {de~Rooij}}\ and\ \bibinfo {author} {\bibfnamefont {E.}~\bibnamefont
  {Kuhl}},\ }\bibfield  {title} {\bibinfo {title} {Physical biology of axonal
  damage},\ }\bibfield  {journal} {\bibinfo  {journal} {Front. Mol. Neurosci.}\
  }\textbf {\bibinfo {volume} {12}},\ \href
  {https://doi.org/10.3389/fncel.2018.00144} {10.3389/fncel.2018.00144}
  (\bibinfo {year} {2018})\BibitemShut {NoStop}%
\bibitem [{\citenamefont {Datar}\ \emph {et~al.}(2019)\citenamefont {Datar},
  \citenamefont {Ameeramja}, \citenamefont {Bhat}, \citenamefont {Srivastava},
  \citenamefont {Mishra}, \citenamefont {Bernal}, \citenamefont {Prost},
  \citenamefont {Callan-Jones},\ and\ \citenamefont {Pullarkat}}]{Datar_2019}%
  \BibitemOpen
  \bibfield  {author} {\bibinfo {author} {\bibfnamefont {A.}~\bibnamefont
  {Datar}}, \bibinfo {author} {\bibfnamefont {J.}~\bibnamefont {Ameeramja}},
  \bibinfo {author} {\bibfnamefont {A.}~\bibnamefont {Bhat}}, \bibinfo {author}
  {\bibfnamefont {R.}~\bibnamefont {Srivastava}}, \bibinfo {author}
  {\bibfnamefont {A.}~\bibnamefont {Mishra}}, \bibinfo {author} {\bibfnamefont
  {R.}~\bibnamefont {Bernal}}, \bibinfo {author} {\bibfnamefont
  {J.}~\bibnamefont {Prost}}, \bibinfo {author} {\bibfnamefont
  {A.}~\bibnamefont {Callan-Jones}},\ and\ \bibinfo {author} {\bibfnamefont
  {P.~A.}\ \bibnamefont {Pullarkat}},\ }\bibfield  {title} {\bibinfo {title}
  {The roles of microtubules and membrane tension in axonal beading,
  retraction, and atrophy},\ }\href {https://doi.org/10.1016/j.bpj.2019.07.046}
  {\bibfield  {journal} {\bibinfo  {journal} {Biophys. J.}\ }\textbf {\bibinfo
  {volume} {117}},\ \bibinfo {pages} {880} (\bibinfo {year}
  {2019})}\BibitemShut {NoStop}%
\bibitem [{\citenamefont {He}\ \emph {et~al.}(2002)\citenamefont {He},
  \citenamefont {Yu},\ and\ \citenamefont {Baas}}]{He_2002}%
  \BibitemOpen
  \bibfield  {author} {\bibinfo {author} {\bibfnamefont {Y.}~\bibnamefont
  {He}}, \bibinfo {author} {\bibfnamefont {W.}~\bibnamefont {Yu}},\ and\
  \bibinfo {author} {\bibfnamefont {P.~W.}\ \bibnamefont {Baas}},\ }\bibfield
  {title} {\bibinfo {title} {Microtubule reconfiguration during axonal
  retraction induced by nitric oxide},\ }\href
  {https://doi.org/10.1523/jneurosci.22-14-05982.2002} {\bibfield  {journal}
  {\bibinfo  {journal} {J. Neurosci.}\ }\textbf {\bibinfo {volume} {22}},\
  \bibinfo {pages} {5982} (\bibinfo {year} {2002})}\BibitemShut {NoStop}%
\bibitem [{\citenamefont {Mora}\ \emph {et~al.}(2010)\citenamefont {Mora},
  \citenamefont {Phou}, \citenamefont {Fromental}, \citenamefont {Pismen},\
  and\ \citenamefont {Pomeau}}]{Mora_2010}%
  \BibitemOpen
  \bibfield  {author} {\bibinfo {author} {\bibfnamefont {S.}~\bibnamefont
  {Mora}}, \bibinfo {author} {\bibfnamefont {T.}~\bibnamefont {Phou}}, \bibinfo
  {author} {\bibfnamefont {J.-M.}\ \bibnamefont {Fromental}}, \bibinfo {author}
  {\bibfnamefont {L.~M.}\ \bibnamefont {Pismen}},\ and\ \bibinfo {author}
  {\bibfnamefont {Y.}~\bibnamefont {Pomeau}},\ }\bibfield  {title} {\bibinfo
  {title} {Capillarity driven instability of a soft solid},\ }\bibfield
  {journal} {\bibinfo  {journal} {Phys. Rev. Lett.}\ }\textbf {\bibinfo
  {volume} {105}},\ \href {https://doi.org/10.1103/physrevlett.105.214301}
  {10.1103/physrevlett.105.214301} (\bibinfo {year} {2010})\BibitemShut
  {NoStop}%
\bibitem [{\citenamefont {Taffetani}\ and\ \citenamefont
  {Ciarletta}(2015)}]{Taffetani_2015}%
  \BibitemOpen
  \bibfield  {author} {\bibinfo {author} {\bibfnamefont {M.}~\bibnamefont
  {Taffetani}}\ and\ \bibinfo {author} {\bibfnamefont {P.}~\bibnamefont
  {Ciarletta}},\ }\bibfield  {title} {\bibinfo {title} {Beading instability in
  soft cylindrical gels with capillary energy: Weakly non-linear analysis and
  numerical simulations},\ }\href {https://doi.org/10.1016/j.jmps.2015.05.002}
  {\bibfield  {journal} {\bibinfo  {journal} {J. Mech. Phys. Solids}\ }\textbf
  {\bibinfo {volume} {81}},\ \bibinfo {pages} {91} (\bibinfo {year}
  {2015})}\BibitemShut {NoStop}%
\bibitem [{\citenamefont {Xuan}\ and\ \citenamefont
  {Biggins}(2017)}]{Xuan_2017}%
  \BibitemOpen
  \bibfield  {author} {\bibinfo {author} {\bibfnamefont {C.}~\bibnamefont
  {Xuan}}\ and\ \bibinfo {author} {\bibfnamefont {J.}~\bibnamefont {Biggins}},\
  }\bibfield  {title} {\bibinfo {title} {Plateau-rayleigh instability in solids
  is a simple phase separation},\ }\bibfield  {journal} {\bibinfo  {journal}
  {Phys. Rev. E}\ }\textbf {\bibinfo {volume} {95}},\ \href
  {https://doi.org/10.1103/physreve.95.053106} {10.1103/physreve.95.053106}
  (\bibinfo {year} {2017})\BibitemShut {NoStop}%
\bibitem [{\citenamefont {Lestringant}\ and\ \citenamefont
  {Audoly}(2020)}]{Lestringant_2020}%
  \BibitemOpen
  \bibfield  {author} {\bibinfo {author} {\bibfnamefont {C.}~\bibnamefont
  {Lestringant}}\ and\ \bibinfo {author} {\bibfnamefont {B.}~\bibnamefont
  {Audoly}},\ }\bibfield  {title} {\bibinfo {title} {A one-dimensional model
  for elasto-capillary necking},\ }\href
  {https://doi.org/10.1098/rspa.2020.0337} {\bibfield  {journal} {\bibinfo
  {journal} {Proc. R. Soc. A}\ }\textbf {\bibinfo {volume} {476}},\ \bibinfo
  {pages} {20200337} (\bibinfo {year} {2020})}\BibitemShut {NoStop}%
\bibitem [{\citenamefont {Giudici}\ and\ \citenamefont
  {Biggins}(2020)}]{Giudici_2020}%
  \BibitemOpen
  \bibfield  {author} {\bibinfo {author} {\bibfnamefont {A.}~\bibnamefont
  {Giudici}}\ and\ \bibinfo {author} {\bibfnamefont {J.~S.}\ \bibnamefont
  {Biggins}},\ }\bibfield  {title} {\bibinfo {title} {Ballooning, bulging, and
  necking: An exact solution for longitudinal phase separation in elastic
  systems near a critical point},\ }\bibfield  {journal} {\bibinfo  {journal}
  {Phys. Rev. E}\ }\textbf {\bibinfo {volume} {102}},\ \href
  {https://doi.org/10.1103/physreve.102.033007} {10.1103/physreve.102.033007}
  (\bibinfo {year} {2020})\BibitemShut {NoStop}%
\bibitem [{\citenamefont {Fu}\ \emph {et~al.}(2021)\citenamefont {Fu},
  \citenamefont {Jin},\ and\ \citenamefont {Goriely}}]{Fu_2021}%
  \BibitemOpen
  \bibfield  {author} {\bibinfo {author} {\bibfnamefont {Y.}~\bibnamefont
  {Fu}}, \bibinfo {author} {\bibfnamefont {L.}~\bibnamefont {Jin}},\ and\
  \bibinfo {author} {\bibfnamefont {A.}~\bibnamefont {Goriely}},\ }\bibfield
  {title} {\bibinfo {title} {Necking, beading, and bulging in soft elastic
  cylinders},\ }\href {https://doi.org/10.1016/j.jmps.2020.104250} {\bibfield
  {journal} {\bibinfo  {journal} {J. Mech. Phys. Solids}\ }\textbf {\bibinfo
  {volume} {147}},\ \bibinfo {pages} {104250} (\bibinfo {year}
  {2021})}\BibitemShut {NoStop}%
\bibitem [{\citenamefont {Pullarkat}\ \emph {et~al.}(2006)\citenamefont
  {Pullarkat}, \citenamefont {Dommersnes}, \citenamefont {Fern{\'{a}}ndez},
  \citenamefont {Joanny},\ and\ \citenamefont {Ott}}]{Pullarkat_2006}%
  \BibitemOpen
  \bibfield  {author} {\bibinfo {author} {\bibfnamefont {P.~A.}\ \bibnamefont
  {Pullarkat}}, \bibinfo {author} {\bibfnamefont {P.}~\bibnamefont
  {Dommersnes}}, \bibinfo {author} {\bibfnamefont {P.}~\bibnamefont
  {Fern{\'{a}}ndez}}, \bibinfo {author} {\bibfnamefont {J.-F.}\ \bibnamefont
  {Joanny}},\ and\ \bibinfo {author} {\bibfnamefont {A.}~\bibnamefont {Ott}},\
  }\bibfield  {title} {\bibinfo {title} {Osmotically driven shape
  transformations in axons},\ }\bibfield  {journal} {\bibinfo  {journal} {Phys.
  Rev. Lett.}\ }\textbf {\bibinfo {volume} {96}},\ \href
  {https://doi.org/10.1103/physrevlett.96.048104}
  {10.1103/physrevlett.96.048104} (\bibinfo {year} {2006})\BibitemShut
  {NoStop}%
\bibitem [{\citenamefont {Letourneau}(2009)}]{Letourneau_2009}%
  \BibitemOpen
  \bibfield  {author} {\bibinfo {author} {\bibfnamefont {P.~C.}\ \bibnamefont
  {Letourneau}},\ }\bibfield  {title} {\bibinfo {title} {Actin in axons: Stable
  scaffolds and dynamic filaments},\ }in\ \href
  {https://doi.org/10.1007/400_2009_15} {\emph {\bibinfo {booktitle} {Results
  and Problems in Cell Differentiation}}}\ (\bibinfo  {publisher} {Springer
  Berlin Heidelberg},\ \bibinfo {year} {2009})\ pp.\ \bibinfo {pages}
  {265--290}\BibitemShut {NoStop}%
\bibitem [{\citenamefont {Liewald}\ \emph {et~al.}(2014)\citenamefont
  {Liewald}, \citenamefont {Miller}, \citenamefont {Logothetis}, \citenamefont
  {Wagner},\ and\ \citenamefont {Schüz}}]{Liewald_2014}%
  \BibitemOpen
  \bibfield  {author} {\bibinfo {author} {\bibfnamefont {D.}~\bibnamefont
  {Liewald}}, \bibinfo {author} {\bibfnamefont {R.}~\bibnamefont {Miller}},
  \bibinfo {author} {\bibfnamefont {N.}~\bibnamefont {Logothetis}}, \bibinfo
  {author} {\bibfnamefont {H.-J.}\ \bibnamefont {Wagner}},\ and\ \bibinfo
  {author} {\bibfnamefont {A.}~\bibnamefont {Schüz}},\ }\bibfield  {title}
  {\bibinfo {title} {Distribution of axon diameters in cortical white matter:
  an electron-microscopic study on three human brains and a macaque},\ }\href
  {https://doi.org/10.1007/s00422-014-0626-2} {\bibfield  {journal} {\bibinfo
  {journal} {Biol. Cybern.}\ }\textbf {\bibinfo {volume} {108}},\ \bibinfo
  {pages} {541} (\bibinfo {year} {2014})}\BibitemShut {NoStop}%
\bibitem [{\citenamefont {Kondaurov}\ and\ \citenamefont
  {Nikitin}(1987)}]{Kondaurov_1987}%
  \BibitemOpen
  \bibfield  {author} {\bibinfo {author} {\bibfnamefont {V.}~\bibnamefont
  {Kondaurov}}\ and\ \bibinfo {author} {\bibfnamefont {L.}~\bibnamefont
  {Nikitin}},\ }\bibfield  {title} {\bibinfo {title} {Finite strains of
  viscoelastic muscle tissue},\ }\href
  {https://doi.org/10.1016/0021-8928(87)90111-0} {\bibfield  {journal}
  {\bibinfo  {journal} {J. Appl. Math. Mech.}\ }\textbf {\bibinfo {volume}
  {51}},\ \bibinfo {pages} {346} (\bibinfo {year} {1987})}\BibitemShut
  {NoStop}%
\bibitem [{\citenamefont {Ambrosi}\ and\ \citenamefont
  {Pezzuto}(2011)}]{Ambrosi_2011}%
  \BibitemOpen
  \bibfield  {author} {\bibinfo {author} {\bibfnamefont {D.}~\bibnamefont
  {Ambrosi}}\ and\ \bibinfo {author} {\bibfnamefont {S.}~\bibnamefont
  {Pezzuto}},\ }\bibfield  {title} {\bibinfo {title} {Active stress vs. active
  strain in mechanobiology: Constitutive issues},\ }\href
  {https://doi.org/10.1007/s10659-011-9351-4} {\bibfield  {journal} {\bibinfo
  {journal} {J. Elasticity}\ }\textbf {\bibinfo {volume} {107}},\ \bibinfo
  {pages} {199} (\bibinfo {year} {2011})}\BibitemShut {NoStop}%
\bibitem [{\citenamefont {Giantesio}\ \emph {et~al.}(2018)\citenamefont
  {Giantesio}, \citenamefont {Musesti},\ and\ \citenamefont
  {Riccobelli}}]{Giantesio_2018}%
  \BibitemOpen
  \bibfield  {author} {\bibinfo {author} {\bibfnamefont {G.}~\bibnamefont
  {Giantesio}}, \bibinfo {author} {\bibfnamefont {A.}~\bibnamefont {Musesti}},\
  and\ \bibinfo {author} {\bibfnamefont {D.}~\bibnamefont {Riccobelli}},\
  }\bibfield  {title} {\bibinfo {title} {A comparison between active strain and
  active stress in transversely isotropic hyperelastic materials},\ }\href
  {https://doi.org/10.1007/s10659-018-9708-z} {\bibfield  {journal} {\bibinfo
  {journal} {J. Elasticity}\ }\textbf {\bibinfo {volume} {137}},\ \bibinfo
  {pages} {63} (\bibinfo {year} {2018})}\BibitemShut {NoStop}%
\bibitem [{\citenamefont {Epstein}(2012)}]{Epstein_2012}%
  \BibitemOpen
  \bibfield  {author} {\bibinfo {author} {\bibfnamefont {M.}~\bibnamefont
  {Epstein}},\ }\href {https://doi.org/10.1002/9781118361016} {\emph {\bibinfo
  {title} {The Elements of Continuum Biomechanics}}}\ (\bibinfo  {publisher}
  {John Wiley {\&} Sons, Ltd},\ \bibinfo {year} {2012})\BibitemShut {NoStop}%
\bibitem [{\citenamefont {Costa}\ \emph {et~al.}(2018)\citenamefont {Costa},
  \citenamefont {Pinto-Costa}, \citenamefont {Sousa},\ and\ \citenamefont
  {Sousa}}]{Costa_2018}%
  \BibitemOpen
  \bibfield  {author} {\bibinfo {author} {\bibfnamefont {A.~R.}\ \bibnamefont
  {Costa}}, \bibinfo {author} {\bibfnamefont {R.}~\bibnamefont {Pinto-Costa}},
  \bibinfo {author} {\bibfnamefont {S.~C.}\ \bibnamefont {Sousa}},\ and\
  \bibinfo {author} {\bibfnamefont {M.~M.}\ \bibnamefont {Sousa}},\ }\bibfield
  {title} {\bibinfo {title} {The regulation of axon diameter: From axonal
  circumferential contractility to activity-dependent axon swelling},\
  }\bibfield  {journal} {\bibinfo  {journal} {Front. Mol. Neurosci.}\ }\textbf
  {\bibinfo {volume} {11}},\ \href {https://doi.org/10.3389/fnmol.2018.00319}
  {10.3389/fnmol.2018.00319} (\bibinfo {year} {2018})\BibitemShut {NoStop}%
\bibitem [{\citenamefont {LoPachin}\ \emph {et~al.}(1991)\citenamefont
  {LoPachin}, \citenamefont {Castiglia},\ and\ \citenamefont
  {Saubermann}}]{LoPachin_1991}%
  \BibitemOpen
  \bibfield  {author} {\bibinfo {author} {\bibfnamefont {R.~M.}\ \bibnamefont
  {LoPachin}}, \bibinfo {author} {\bibfnamefont {C.~M.}\ \bibnamefont
  {Castiglia}},\ and\ \bibinfo {author} {\bibfnamefont {A.~J.}\ \bibnamefont
  {Saubermann}},\ }\bibfield  {title} {\bibinfo {title} {Elemental composition
  and water content of myelinated axons and glial cells in rat central nervous
  system},\ }\href {https://doi.org/10.1016/0006-8993(91)90465-8} {\bibfield
  {journal} {\bibinfo  {journal} {Brain Res.}\ }\textbf {\bibinfo {volume}
  {549}},\ \bibinfo {pages} {253} (\bibinfo {year} {1991})}\BibitemShut
  {NoStop}%
\bibitem [{\citenamefont {Ogden}(1997)}]{ogden1997non}%
  \BibitemOpen
  \bibfield  {author} {\bibinfo {author} {\bibfnamefont {R.~W.}\ \bibnamefont
  {Ogden}},\ }\href@noop {} {\emph {\bibinfo {title} {Non-linear elastic
  deformations}}}\ (\bibinfo  {publisher} {Courier Corporation},\ \bibinfo
  {year} {1997})\BibitemShut {NoStop}%
\bibitem [{\citenamefont {Bigoni}\ and\ \citenamefont
  {Gei}(2001)}]{Bigoni_2001}%
  \BibitemOpen
  \bibfield  {author} {\bibinfo {author} {\bibfnamefont {D.}~\bibnamefont
  {Bigoni}}\ and\ \bibinfo {author} {\bibfnamefont {M.}~\bibnamefont {Gei}},\
  }\bibfield  {title} {\bibinfo {title} {Bifurcations of a coated, elastic
  cylinder},\ }\href {https://doi.org/10.1016/s0020-7683(00)00322-x} {\bibfield
   {journal} {\bibinfo  {journal} {Int. J. Solids Struct.}\ }\textbf {\bibinfo
  {volume} {38}},\ \bibinfo {pages} {5117} (\bibinfo {year}
  {2001})}\BibitemShut {NoStop}%
\bibitem [{\citenamefont {Stroh}(1962)}]{Stroh_1962}%
  \BibitemOpen
  \bibfield  {author} {\bibinfo {author} {\bibfnamefont {A.~N.}\ \bibnamefont
  {Stroh}},\ }\bibfield  {title} {\bibinfo {title} {Steady state problems in
  anisotropic elasticity},\ }\href@noop {} {\bibfield  {journal} {\bibinfo
  {journal} {Journal of Mathematics and Physics}\ }\textbf {\bibinfo {volume}
  {41}},\ \bibinfo {pages} {77} (\bibinfo {year} {1962})}\BibitemShut {NoStop}%
\bibitem [{\citenamefont {Fu}(2007)}]{Fu_2007}%
  \BibitemOpen
  \bibfield  {author} {\bibinfo {author} {\bibfnamefont {Y.}~\bibnamefont
  {Fu}},\ }\bibfield  {title} {\bibinfo {title} {Hamiltonian interpretation of
  the stroh formalism in anisotropic elasticity},\ }\href
  {https://doi.org/10.1098/rspa.2007.0093} {\bibfield  {journal} {\bibinfo
  {journal} {Proc. R. Soc. A}\ }\textbf {\bibinfo {volume} {463}},\ \bibinfo
  {pages} {3073} (\bibinfo {year} {2007})}\BibitemShut {NoStop}%
\bibitem [{\citenamefont {Norris}\ and\ \citenamefont
  {Shuvalov}(2010)}]{Norris_2010}%
  \BibitemOpen
  \bibfield  {author} {\bibinfo {author} {\bibfnamefont {A.~N.}\ \bibnamefont
  {Norris}}\ and\ \bibinfo {author} {\bibfnamefont {A.~L.}\ \bibnamefont
  {Shuvalov}},\ }\bibfield  {title} {\bibinfo {title} {Wave impedance matrices
  for cylindrically anisotropic radially inhomogeneous elastic solids},\
  }\href@noop {} {\bibfield  {journal} {\bibinfo  {journal} {Quart. J. Mech.
  Appl. Math.}\ }\textbf {\bibinfo {volume} {63}},\ \bibinfo {pages} {401}
  (\bibinfo {year} {2010})}\BibitemShut {NoStop}%
\bibitem [{\citenamefont {Cao}\ \emph {et~al.}(2012)\citenamefont {Cao},
  \citenamefont {Li},\ and\ \citenamefont {Feng}}]{Cao_2012}%
  \BibitemOpen
  \bibfield  {author} {\bibinfo {author} {\bibfnamefont {Y.-P.}\ \bibnamefont
  {Cao}}, \bibinfo {author} {\bibfnamefont {B.}~\bibnamefont {Li}},\ and\
  \bibinfo {author} {\bibfnamefont {X.-Q.}\ \bibnamefont {Feng}},\ }\bibfield
  {title} {\bibinfo {title} {Surface wrinkling and folding of
  core{\textendash}shell soft cylinders},\ }\href
  {https://doi.org/10.1039/c1sm06354e} {\bibfield  {journal} {\bibinfo
  {journal} {Soft Matter}\ }\textbf {\bibinfo {volume} {8}},\ \bibinfo {pages}
  {556} (\bibinfo {year} {2012})}\BibitemShut {NoStop}%
\bibitem [{\citenamefont {Ciarletta}\ \emph {et~al.}(2014)\citenamefont
  {Ciarletta}, \citenamefont {Balbi},\ and\ \citenamefont
  {Kuhl}}]{ciarletta2014pattern}%
  \BibitemOpen
  \bibfield  {author} {\bibinfo {author} {\bibfnamefont {P.}~\bibnamefont
  {Ciarletta}}, \bibinfo {author} {\bibfnamefont {V.}~\bibnamefont {Balbi}},\
  and\ \bibinfo {author} {\bibfnamefont {E.}~\bibnamefont {Kuhl}},\ }\bibfield
  {title} {\bibinfo {title} {Pattern selection in growing tubular tissues},\
  }\href@noop {} {\bibfield  {journal} {\bibinfo  {journal} {Phys. Rev. Lett.}\
  }\textbf {\bibinfo {volume} {113}},\ \bibinfo {pages} {248101} (\bibinfo
  {year} {2014})}\BibitemShut {NoStop}%
\bibitem [{\citenamefont {Dervaux}\ and\ \citenamefont
  {Amar}(2011)}]{Dervaux_2011}%
  \BibitemOpen
  \bibfield  {author} {\bibinfo {author} {\bibfnamefont {J.}~\bibnamefont
  {Dervaux}}\ and\ \bibinfo {author} {\bibfnamefont {M.~B.}\ \bibnamefont
  {Amar}},\ }\bibfield  {title} {\bibinfo {title} {Buckling condensation in
  constrained growth},\ }\href {https://doi.org/10.1016/j.jmps.2010.12.015}
  {\bibfield  {journal} {\bibinfo  {journal} {J. Mech. Phys. Solids}\ }\textbf
  {\bibinfo {volume} {59}},\ \bibinfo {pages} {538} (\bibinfo {year}
  {2011})}\BibitemShut {NoStop}%
\bibitem [{\citenamefont {Ciarletta}\ \emph {et~al.}(2016)\citenamefont
  {Ciarletta}, \citenamefont {Destrade}, \citenamefont {Gower},\ and\
  \citenamefont {Taffetani}}]{ciarletta2016morphology}%
  \BibitemOpen
  \bibfield  {author} {\bibinfo {author} {\bibfnamefont {P.}~\bibnamefont
  {Ciarletta}}, \bibinfo {author} {\bibfnamefont {M.}~\bibnamefont {Destrade}},
  \bibinfo {author} {\bibfnamefont {A.~L.}\ \bibnamefont {Gower}},\ and\
  \bibinfo {author} {\bibfnamefont {M.}~\bibnamefont {Taffetani}},\ }\bibfield
  {title} {\bibinfo {title} {Morphology of residually stressed tubular tissues:
  Beyond the elastic multiplicative decomposition},\ }\href@noop {} {\bibfield
  {journal} {\bibinfo  {journal} {J. Mech. Phys. Solids}\ }\textbf {\bibinfo
  {volume} {90}},\ \bibinfo {pages} {242} (\bibinfo {year} {2016})}\BibitemShut
  {NoStop}%
\bibitem [{\citenamefont {Du}\ \emph {et~al.}(2019)\citenamefont {Du},
  \citenamefont {Lü}, \citenamefont {Destrade},\ and\ \citenamefont
  {Chen}}]{Du_2019}%
  \BibitemOpen
  \bibfield  {author} {\bibinfo {author} {\bibfnamefont {Y.}~\bibnamefont
  {Du}}, \bibinfo {author} {\bibfnamefont {C.}~\bibnamefont {Lü}}, \bibinfo
  {author} {\bibfnamefont {M.}~\bibnamefont {Destrade}},\ and\ \bibinfo
  {author} {\bibfnamefont {W.}~\bibnamefont {Chen}},\ }\bibfield  {title}
  {\bibinfo {title} {Influence of initial residual stress on growth and pattern
  creation for a layered aorta},\ }\bibfield  {journal} {\bibinfo  {journal}
  {Sci. Rep.}\ }\textbf {\bibinfo {volume} {9}},\ \href
  {https://doi.org/10.1038/s41598-019-44694-2} {10.1038/s41598-019-44694-2}
  (\bibinfo {year} {2019})\BibitemShut {NoStop}%
\bibitem [{\citenamefont {Boffi}\ \emph {et~al.}(2013)\citenamefont {Boffi},
  \citenamefont {Brezzi},\ and\ \citenamefont {Fortin}}]{boffi2013mixed}%
  \BibitemOpen
  \bibfield  {author} {\bibinfo {author} {\bibfnamefont {D.}~\bibnamefont
  {Boffi}}, \bibinfo {author} {\bibfnamefont {F.}~\bibnamefont {Brezzi}},\ and\
  \bibinfo {author} {\bibfnamefont {M.}~\bibnamefont {Fortin}},\ }\href@noop {}
  {\emph {\bibinfo {title} {Mixed finite element methods and applications}}},\
  Vol.~\bibinfo {volume} {44}\ (\bibinfo  {publisher} {Springer},\ \bibinfo
  {year} {2013})\BibitemShut {NoStop}%
\bibitem [{\citenamefont {Riccobelli}\ \emph {et~al.}(2021)\citenamefont
  {Riccobelli}, \citenamefont {Noselli},\ and\ \citenamefont
  {DeSimone}}]{Riccobelli_2021}%
  \BibitemOpen
  \bibfield  {author} {\bibinfo {author} {\bibfnamefont {D.}~\bibnamefont
  {Riccobelli}}, \bibinfo {author} {\bibfnamefont {G.}~\bibnamefont
  {Noselli}},\ and\ \bibinfo {author} {\bibfnamefont {A.}~\bibnamefont
  {DeSimone}},\ }\bibfield  {title} {\bibinfo {title} {Rods coiling about a
  rigid constraint: helices and perversions},\ }\href
  {https://doi.org/10.1098/rspa.2020.0817} {\bibfield  {journal} {\bibinfo
  {journal} {Proc. R. Soc. A}\ }\textbf {\bibinfo {volume} {477}},\ \bibinfo
  {pages} {20200817} (\bibinfo {year} {2021})}\BibitemShut {NoStop}%
\bibitem [{Note1()}]{Note1}%
  \BibitemOpen
  \bibinfo {note} {See Supplemental Material for videos showing the
  post-buckling evolution predicted by the numerical simulations.}\BibitemShut
  {Stop}%
\bibitem [{\citenamefont {Balbi}\ \emph {et~al.}(2020)\citenamefont {Balbi},
  \citenamefont {Destrade},\ and\ \citenamefont {Goriely}}]{Balbi_2020}%
  \BibitemOpen
  \bibfield  {author} {\bibinfo {author} {\bibfnamefont {V.}~\bibnamefont
  {Balbi}}, \bibinfo {author} {\bibfnamefont {M.}~\bibnamefont {Destrade}},\
  and\ \bibinfo {author} {\bibfnamefont {A.}~\bibnamefont {Goriely}},\
  }\bibfield  {title} {\bibinfo {title} {Mechanics of human brain organoids},\
  }\bibfield  {journal} {\bibinfo  {journal} {Phys. Rev. E}\ }\textbf {\bibinfo
  {volume} {101}},\ \href {https://doi.org/10.1103/physreve.101.022403}
  {10.1103/physreve.101.022403} (\bibinfo {year} {2020})\BibitemShut {NoStop}%
\bibitem [{\citenamefont {Riccobelli}\ and\ \citenamefont
  {Bevilacqua}(2020)}]{Riccobelli_2020}%
  \BibitemOpen
  \bibfield  {author} {\bibinfo {author} {\bibfnamefont {D.}~\bibnamefont
  {Riccobelli}}\ and\ \bibinfo {author} {\bibfnamefont {G.}~\bibnamefont
  {Bevilacqua}},\ }\bibfield  {title} {\bibinfo {title} {Surface tension
  controls the onset of gyrification in brain organoids},\ }\href
  {https://doi.org/10.1016/j.jmps.2019.103745} {\bibfield  {journal} {\bibinfo
  {journal} {J. Mech. Phys. Solids}\ }\textbf {\bibinfo {volume} {134}},\
  \bibinfo {pages} {103745} (\bibinfo {year} {2020})}\BibitemShut {NoStop}%
\end{thebibliography}
\end{document}